\documentclass[
twocolumn,
superscriptaddress,
showpacs,
amsmath,amssymb,
 aps,
prl
]{revtex4-2}

\usepackage{mathtools}
\usepackage{amsmath}
\usepackage{amssymb}
\usepackage{bbm}
\usepackage{bm}
\usepackage{graphicx}
\usepackage{xcolor}
\usepackage{dcolumn}
\usepackage{bm}
\usepackage{float}
\usepackage{enumitem}

\usepackage{algorithm}
\usepackage{algpseudocode}

\usepackage{subfigure}
\usepackage[colorlinks]{hyperref}
\usepackage{dsfont}
\usepackage{braket}
\usepackage{units}
\usepackage{placeins}

\graphicspath{{./},{results/},{results/dmft/},{results/test_NB_6/}}

\newcommand{\matK}[1]{\underline{#1}}

\DeclareMathOperator*{\RRe}{Re}
\DeclareMathOperator*{\IIm}{Im}

\newcommand{\phys}{_\text{phys}}
\newcommand{\aux}{_\text{aux}}
\newcommand{\eps}{\varepsilon}
\newcommand{\meps}{\matK{\eps}}
\newcommand{\mmu}{\matK{\mu}}
\newcommand{\fde}{\delta}
\newcommand{\funcderiv}{\frac{\fde \matK{\Sigma}[\matK{\Delta}\phys]}{\fde \matK{\Delta}\phys}}
\newcommand{\beq}{\begin{equation}}
\newcommand{\eeq}{\end{equation}}
\newcommand{\vv}{\bm}
\newcommand{\des}{_\text{des}}
\usepackage{xspace}

\newcommand{\FI}{Functional interpolation\xspace}
\newcommand{\fI}{functional interpolation\xspace}

\begin{document}

\title{\FI expansion for nonequilibrium correlated impurities}

\author{Daniel Werner}
\email[]{daniel.werner96@posteo.at}
\affiliation{Institute of Theoretical and Computational Physics, Graz University of Technology, 8010 Graz, Austria}
\author{Enrico Arrigoni}
\email[]{arrigoni@tugraz.at}
\affiliation{Institute of Theoretical and Computational Physics, Graz University of Technology, 8010 Graz, Austria}

\begin{abstract}
	We present a \fI approach within the auxiliary master equation framework to efficiently and accurately solve correlated impurity problems in nonequilibrium dynamical mean-field theory (DMFT). By leveraging a near-exact auxiliary bath representation, the method estimates corrections via interpolation over a few bath realisations, significantly reducing computational cost and increasing accuracy. We illustrate the approach on the Anderson impurity model and on
	the Hubbard model within DMFT, capturing equilibrium and long-lived photodoped states.

\end{abstract}

\date{\today}

\maketitle

\paragraph{Introduction}
Advances in controlling quantum materials, ranging from ultrafast laser spectroscopy,~\cite{iw.on.03,ca.de.04,pe.lo.06,fa.to.11} solid state nanocience,~\cite{zu.fa.04,bo.gr.05} and ultracold quantum gases~\cite{ra.sa.97,ja.br.98,gr.ma.02,ha.br.08} have sparked growing interest in nonequilibrium strongly correlated systems.
These developments drive theoretical challenges, including understanding nonequilibrium quantum phase transitions,~\cite{mi.ta.06} thermalization, and decoherence~\cite{caza.06,ca.ca.07,ri.du.08,le.ch.87}.
In this context, the theoretical description of strongly correlated quantum systems out of equilibrium remains an exciting challenge in modern theoretical physics.

The extension of dynamical mean-field theory (DMFT)~\cite{ge.ko.96,me.vo.89} to nonequilibrium~\cite{sc.mo.02u,fr.tu.06,free.08,jo.fr.08,ec.ko.09,okam.07,ao.ts.14} provides a key framework for addressing these systems. DMFT in the time domain has been applied to simulate
a variety of phenomena and systems ranging from highly excited states in photoexcited materials to
ultracold atoms or Mott insulators under strong DC fields (see, e.g. Ref.\cite{ao.ts.14}).
The main bottleneck of DMFT, the solution of a many-body quantum impurity, is a long-standing problem for nonequilibrium systems.
Exact quantum Monte Carlo (QMC)~\cite{gu.mi.11,we.ok.09} or matrix-product states~\cite{wh.fe.04,da.ko.04}
calculations are limited to short time so that mostly low-order hybridization expansions, such as the non-crossing approximation are adopted in practice.
When restricting to the steady state, advances have been achieved, among others, via the QMC inchworm~\cite{er.gu.23}, scattering-states approaches~\cite{me.an.06,ande.08},
or the auxiliary master equation (AMEA)~\cite{ar.kn.13,do.nu.14} algorithms.

Steady-state algorithms provide a
complement to real-time simulations as they can be used to
address the slowly time-evolving regime occurring, for instance in photoexcited Mott insulators after an initial prethermalisation phase~\cite{da.li.21,ro.ra.08,se.pe.10,mu.ta.22,le.pr.13}.
While this regime cannot be reliably addressed by numerically exact real-time approaches,
one can adopt a time-local ansatzt for the fermionic distribution
function of a reference steady-state system~\cite{ku.er.24} which then evolves in time following a generalized quantum Boltzmann scheme~\cite{pi.li.21}.

It is therefore crucial
to develop nonperturbative, accurate methods
that remain computationally efficient enough to be used as quantum impurity solvers in DMFT self-consistent calculations.

The approach presented in this Letter exploits the freedom to replace
the ``physical'' fermionic bath of the impurity problem that needs to be solved, e.g. at each DMFT step,
with a so-called ``auxiliary'' one consisting of a small number $N_\text{b}$ (typically $6$ to $8$) of bath sites connected to a Markovian environment which then can be solved exactly by numerical methods.
Since the ``residual'' distance (termed $\meps$ below) between the
hybridization functions of the two baths is exponentially small in $N_\text{b}$~\cite{do.so.17}
a substantial improvement by including perturbative corrections in $\meps$ can be obtained, see Ref.~\cite{ch.co.19} for a similar idea.
However, for values of $N_\text{b}\gtrsim 6$ for which $\meps$ is small enough, this expansion becomes prohibitive and computationally quite expensive.
The main idea presented here consists in
adopting
a \FI (FI) approach which amounts in estimating the
$\meps$ linear corrections by
a solution of the impurity problem for a few (typically two to four)
different suitably chosen sets of bath parameters.

After testing the approach for the Anderson impurity model in and out of equilibium,
we address, within DMFT, the Hubbard model on both sides of the equilibrium Mott phase, as well as the long-lived nonequilibrium photodoped state.

\paragraph{\label{method} Model and Method}

A fermionic single-impurity problem is fully characterized by its impurity hamiltonian
$H_\text{imp}$ typically describing electrons on a single orbital with an onsite interaction $U$.
The coupling of this orbital to a noninteracting fermionic bath is completely characterized
by its hybridization function $\matK{\Delta}_\text{phys}(\omega)$, which in a nonequilibrium steady state consists of a $2\times2$ matrix in Keldysh space~\cite{ha.ja,schw.61,keld.65,kad.baym,ra.sm.86}~\footnote{Here and in the following, we use underscore $\matK{X}$ to denote this Keldysh structure containing the retarded $X^R$, Keldysh $X^K$ and advanced $X^A$ components~\cite{ha.ja,ra.sm.86}.
	$X$ can be any two-point function, such as $\Delta$, $G$, or $\Sigma$, or their differences, such as the quantities $\eps$ or $\mu$ below.
	In addition, we shall omit the argument $\omega$ unless neccessary.}.

Within AMEA~\cite{ar.kn.13,do.nu.14} this ``physical'' hybridization function
$\matK{\Delta}\phys(\omega)$
is approximated by
the hybridization function
\begin{equation}
	\label{deltaeps}
	\matK{\Delta}\aux(\omega) = \matK{\Delta}\phys(\omega) + \meps(\omega)
\end{equation}
of an ``auxiliary''

bath consisting of a finite number $N_\text{b}$ of bath sites connected to a Markovian environment,
whose dynamics is described by the Lindblad equation, see ~\cite{do.nu.14,ar.kn.13}.
For a sufficiently small $N_\text{b}$, the impurity problem connected to the auxiliary bath
can be solved exactly, for example, by Lanczos/Arnoldi-like methods~\cite{do.nu.14,we.lo.23}, by matrix-product states (MPS)~\cite{do.ga.15} or by stochastic wave function approaches~\cite{so.fu.19}. In principle, in the $\meps\to 0$ limit the computed {\em interacting} impurity Green's function of this auxiliary system approaches the physical one for any value of $U$.
A clever choice of the (Lindblad) parameters~\cite{do.so.17} of the auxiliary bath obtained, e.g., by fitting
$\matK{\Delta}\aux$ to $\matK{\Delta}\phys$ makes $|\meps|$ to vanish exponentially in terms of $N_\text{b}$~\cite{do.so.17}. Here $|\meps|$ is some measure of the norm of the
deviation $\meps$ integrated over frequencies possibly
with some weight factor
$W(\omega)$ to emphasize some physical relevant frequency regions.

Unfortunately, since one must consider the full space of many-body density matrices,
the maximum $N_\text{b}$ that can be achieved by Lanczos-like methods~\footnote{Combined by a configuraion interaction (CI) approach~\cite{we.lo.23}}
is restricted to $N_\text{b}\approx 8-10$.
While MPS approaches can address larger systems,
the geometry restriction makes the fit less efficient~\cite{do.so.17}.
We, therefore, propose to
improve the accuracy
by including corrections of $O(\meps^1)$.
In principle, this is the spirit of hybridization expansion methods
for example the non-crossing approximation
(NCA)~\cite{ke.ki.70,cole.84,ec.we.10}, which
corresponds, formally, to taking $\meps=\matK \Delta\phys$ in \eqref{deltaeps}.
However, since $\meps$ is not small in this case,
only higher orders, provide reliable results. For these one has to resort to,
e.g., Quantum Monte
Carlo~\cite{we.ok.09,co.gu.15,er.gu.23} or, as shown recently, to tensor
trains~\cite{nu.je.22,ecks.24u}.

In Ref.~\cite{ch.go.19} it was suggested to carry out a dual-fermion expansion around a ``better'' solution with a
$\matK \Delta\aux$ closer to the physical one obtained with
an auxiliary bath with $N_\text{b}=2$.
While the results of the expansion display a significant improvement with respect to the one of the plain auxiliary system~\footnote{In Ref.~\cite{ch.go.19} the auxiliary system is referred to as reference system.}, an extension to larger $N_\text{b}$ is quite prohibitive.
This is due to the fact that one needs to evaluate a three-point vertex, which is numerically complicated in Lanczos/Arnoldi.
On the other hand, only values of $N_\text{b} \gtrsim 6$ guarantee that $\meps$ is sufficiently small so that a first order expansion is justified.
Here,
we follow a different path and estimate the $O(\meps^1)$ term numerically
by solving the nonequilibrium impurity problem for
a batch
of $\meps$.
We exploit the
fact that, for fixed $H_\text{imp}$
the steady-state self energy $\matK{\Sigma}$ is a functional of
$\matK{\Delta}$ only.
Therefore, to first order in $\meps$ we can write
\beq
\label{fi}
\matK{\Sigma}[\matK{\Delta}\phys + \meps] = \matK{\Sigma}[\matK{\Delta}\phys] +
\funcderiv
\odot {\meps}
+ O(\meps)^2.
\eeq
Strictly speaking, the independent functions in the argument of $\matK{\Sigma}$
are just $\IIm \Delta^\text{R}(\omega)$ and
$\IIm \Delta^\text{K}(\omega)$, as $\RRe \Delta^\text{R}$ is fixed by Kramers-Kr\"onig
and $\RRe \Delta^\text{K}=0$. Therefore, we will understand the functional dependence, and corresponding derivative,
with respect to these functions only.
Similarly,  $\odot$ indicates a convolution, i.e. a sum over these two components and an integral over $\omega$.
An equation like \eqref{fi} is obviously valid for the Green's function as well. Here, we will limit the discussion to the self-energy.

To start with,
let us first hypothetically assume that,
in addition to \eqref{deltaeps},
one could compute the self energy for a different ``desired'' hybridization
\beq
\label{des}
\matK{\Delta}\des = \matK{\Delta}\phys + k\ \meps
\eeq
with some real number $k$ away from 1.
Then, by applying \eqref{fi} for $\meps$ replaced with $ k\cdot \meps$ one can cancel the first order errors exactly yielding
\beq
\label{cancel}
\matK{\Sigma}[\matK{\Delta}\phys] = \frac{1}{1 - k} \left ( \matK{\Sigma}[\matK{\Delta}_\text{phys} + k \ \meps] - k \ \matK{\Sigma}[\matK{\Delta}_\text{phys} + \meps] \right )
+ O(\meps^2),
\eeq
For example, for $k=-1$ this is the average between the two self-energies at
$\matK{\Delta}\phys \pm \meps$.

The problem is that, in general, $\matK{\Delta}\des$ will not be representable within
an auxiliary bath with finite $N_\text{b}$ in the sense that there is no set of Lindblad parameters producing it.

An alternative ``bottom-up'' approach consists in generalizing \eqref{cancel} to a
set of
$n_{\max}$ different
suitably chosen auxiliary baths with the same $N_\text{b}$ but with different parameters.
The deviations $\meps_n$ of their hybridization functions
from the physical one

\beq
\label{epsn}
\meps_n= \matK{\Delta}\phys-{\matK{\Delta}\aux}_n \quad\quad n=1,\cdots,n_{\max} \;.
\eeq
must remain small, i.e. of the same order of magnitude.

One can then determine coefficients $\alpha_n$
such that the quantity
\beq
\label{mu}
\mmu \equiv \sum_{n=1}^{n_{\max}} \alpha_n \ \meps_n
\quad\text{with} \quad \sum_{n=1}^{n_{\max}} \alpha_n=1
\eeq
is much smaller than the $\meps_n$.
Here, $\mmu(\omega)$ controls the ``residual'' first-oder error
in an
improved approximation for the self-energy of the physical system
\beq
\label{sigmaimp}
\Sigma[\matK{\Delta}\phys] =
\sum_{n=1}^{n_{\max}} \alpha_n \ \matK{\Sigma}[\matK{\Delta}\phys + \meps_n]
+O(\mmu)
+ O(\meps^2)\;.
\eeq
The optimal coefficients
$\alpha_n$ should be, thus, chosen such that

\begin{equation}
	|\mmu|^2 = \sum_{m,n=1}^{n_{\max}} \alpha_n \alpha_m A_{mn} \quad\quad
	A_{mn} \equiv \meps_n \odot \meps_m
\end{equation}
is minimal. The constraint
in \eqref{mu}
can be addressed via a Lagrange multiplier $\lambda$ leading to the optimal values
\beq
\label{alpha}
\vv \alpha = \lambda \vv A^{-1} \cdot \vv u \quad\quad
\lambda = \left(\vv u^T \cdot \vv A^{-1} \vv u \right)^{-1}
\eeq
with $\vv u^T = (1,1,1,\dots)$.

We have tried two different procedures
to select the set of
${\matK{\Delta}\aux}_n$in \eqref{epsn}, or, equivalently
$\meps_n$.
The first one is {\em deterministic} and consists in the following steps:
We start by finding the (Lindblad) parameters of the auxiliary bath that give the best fit to
${\matK{\Delta}\phys}$. The correspondig
hybridization function is assigned to
${\matK{\Delta}\aux}_m$
with $m=1$. \\
(i) For a given ${\matK{\Delta}\aux}_m$ we extract
the corresponding $\meps_m$ from \eqref{epsn} and construct the corresponding
$\mmu$
from \eqref{mu} and \eqref{alpha} with $n_{\max}=m$.\\
(ii) We use this $\mmu$, which we name $\mmu_m$,
to
determine a new ${\matK{\Delta}\aux}_{m+1}$
by fitting the Lindblad parameters to a
``desired'' ${\matK{\Delta}\des}_{m+1}=\matK{\Delta}\phys + k \ \matK{\mu}_m$ (cf. \eqref{des}).
Here, $k$ is some number~\footnote{$k$ can, in principle, change as a function of $n$.} sufficiently away from $1$.
This choice is motivated by the fact that if one could fit
$\matK{\Delta}\des$ exactly, the first order error would be canceled, similarly to \eqref{cancel}.
With the new ${\matK{\Delta}\aux}_{m+1}$ we go back to step (i).\\
Steps (i) to (ii) should be repeated up to
a maximum value $N_\text{T}$ for $m$
or until
the condition number~\footnote{The condition number is a measure of how ill-conditioned the matrix inversion is, see. e.g.
	\url{https://numpy.org/doc/2.1/reference/generated/numpy.linalg.cond.html}
}
of the matrix $A$ exceeds a given value.
Finally, the
improved self-energy is obtained via \eqref{sigmaimp}.

A modification of this procedure consists in starting with two (or more) different initial ${\matK{\Delta}\aux}_{m=1}$ obtained by different fits, i.e. using
different weight functions $W(\omega)$.
Then the corresponding $\meps_m$ are combined together in
\eqref{sigmaimp}.
Alternatively, starting with the same ${\matK{\Delta}\aux}_{m=1}$ as for the previous algorithm,
one can adopt a random procedure for determining the
$\meps_m$ for $m> 1$. This is achieved by replacing the $N_\text{T}$ ``desired'' hybridization functions of step (ii) in the previous procedure by
random functions sufficiently close to $\matK{\Delta}\phys(\omega)$.
~\footnote{Notice that ${\matK{\Delta}\des}_m$ may become unphysical, for example non-causal. This is, however, not an issue, as
	the fitting procedure ensures that
	${\matK{\Delta}\aux}_m$ remains physical.}
Since this is a random procedure, it may be convenient to repeat it
a certain number of times $N_\text{av}$ and average the corresponding results for the self energy.

\paragraph{\label{results} Results}
Since we have a certain freedom in choosing the details of the procedure, such as the value of $k$ and $N_\text{T}$ or whether to use the deterministic or the random version, it is convenient to first test different options in order to find out the most efficient and accurate one.
We do this by applying the approach for  numerically ``cheap'' auxiliary baths with $N_\text{b}=4$ and compare the self-energy obtained in this way with a reference evaluated with the
plain~\footnote{By ``plain'' we mean without the FI improvement, correpsonding to $N_\text{T}=1$.} $N_\text{b}=8$ one which is expected to be more accurate than the $N_\text{b}=4$ case.
The benchmark reported in the supplemental material~\cite{suppmat} suggests that the best accuracy is obtained with the deterministic algorithm with $k=2$ and $N_\text{T}=2$. This is interesting, as it shows that we can substantially increase the accuracy of the impurity solver by just carrying out two Lindblad many-body calculations.
In particular, as we will see below, a many-body calculation with $N_\text{b}=6$ using the present FI approach yields an accuracy comparable or better than a plain $N_\text{b}=8$ one, which is approximately
$6$ times as costly. In addition, the two many-body calculations required by the FI
approach can be done in parallel.
This is particularly convenient when using it as an impurity solver for Dynamical Mean-Field Theory (DMFT)

\begin{figure}
	\centering
	\includegraphics[scale=0.35]{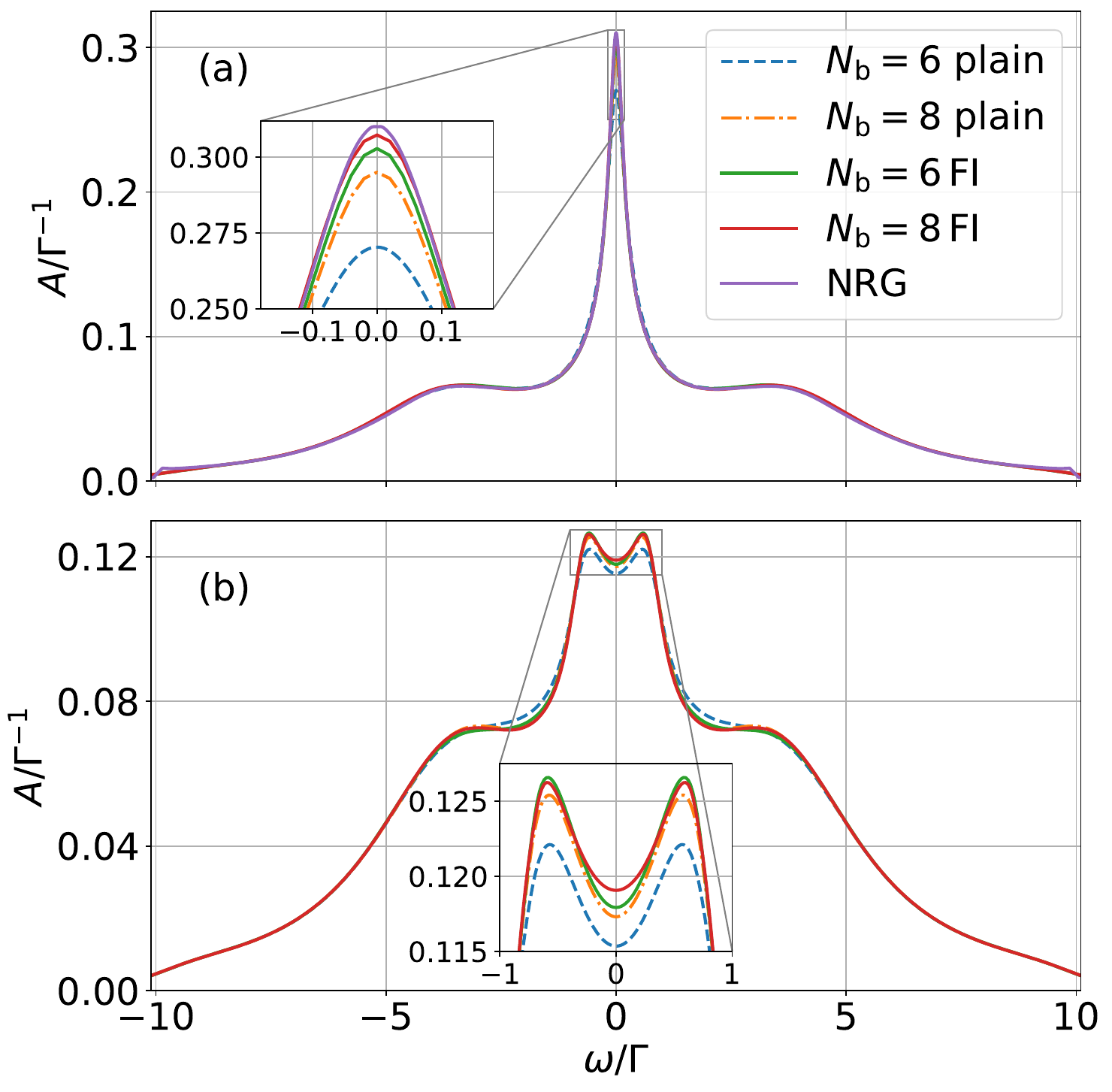}
	\caption{
		Spectral function of the equilibrium (a) Anderson impurity model for
		$U/\Gamma = 6$, $-\IIm \Delta^\text{R}(\omega = 0)/\Gamma = 1$, $\varepsilon_{\text{imp}} = -U/2$, and $T/\Gamma = 0.02$ obtained by
		plain AMEA
		with $N_\text{b}=6$ and $N_\text{b}=8$
		and improved results obtained by the FI approach with $k=2$ and $N_\text{T}=2$ for the two cases.
		(b) shows the same results for the case of a finite bias voltage $V/\Gamma=1.5$.
		The equilibrium ($V/\Gamma=0$) results are compared with NRG.
	}
	\label{gimp}
\end{figure}

We start by
computing
the spectral function for an Anderson impurity in and out of equilibrium.
Unless otherwise specified, we consider a bath with a flat density of states (DOS) of
width $10 \Gamma$~\footnote{The edges of the DOS are smoothed with a ``fictitious'' temperature $0.5\Gamma$.}
with the following parameters
$U/\Gamma = 6$, $-\IIm \Delta^\text{R}(\omega = 0)/\Gamma=1$, $\varepsilon_{\text{imp}} = -U/2$.
In Fig.~\ref{gimp} we display the spectral function of the impurity coupled to (a) a bath in equilibrium at temperature $T/\Gamma=0.02$
and (b) to two baths with a bias voltage $V/\Gamma=1.5$. As one can see, the results with $N_\text{b}=6$ augmented with the FI approach is in some cases even more accurate than the plain $N_\text{b}=8$, as the comparison with Numerical Renormalisation Group (NRG) shows.
FI combined with $N_\text{b}=8$ provides an even better accuracy.
Of course, in equilibrium the method cannot compete with NRG, but it has the advantage that it can be straightforwadly extended to nonequilibrium steady states.

\begin{figure}
	\includegraphics[scale=0.42]{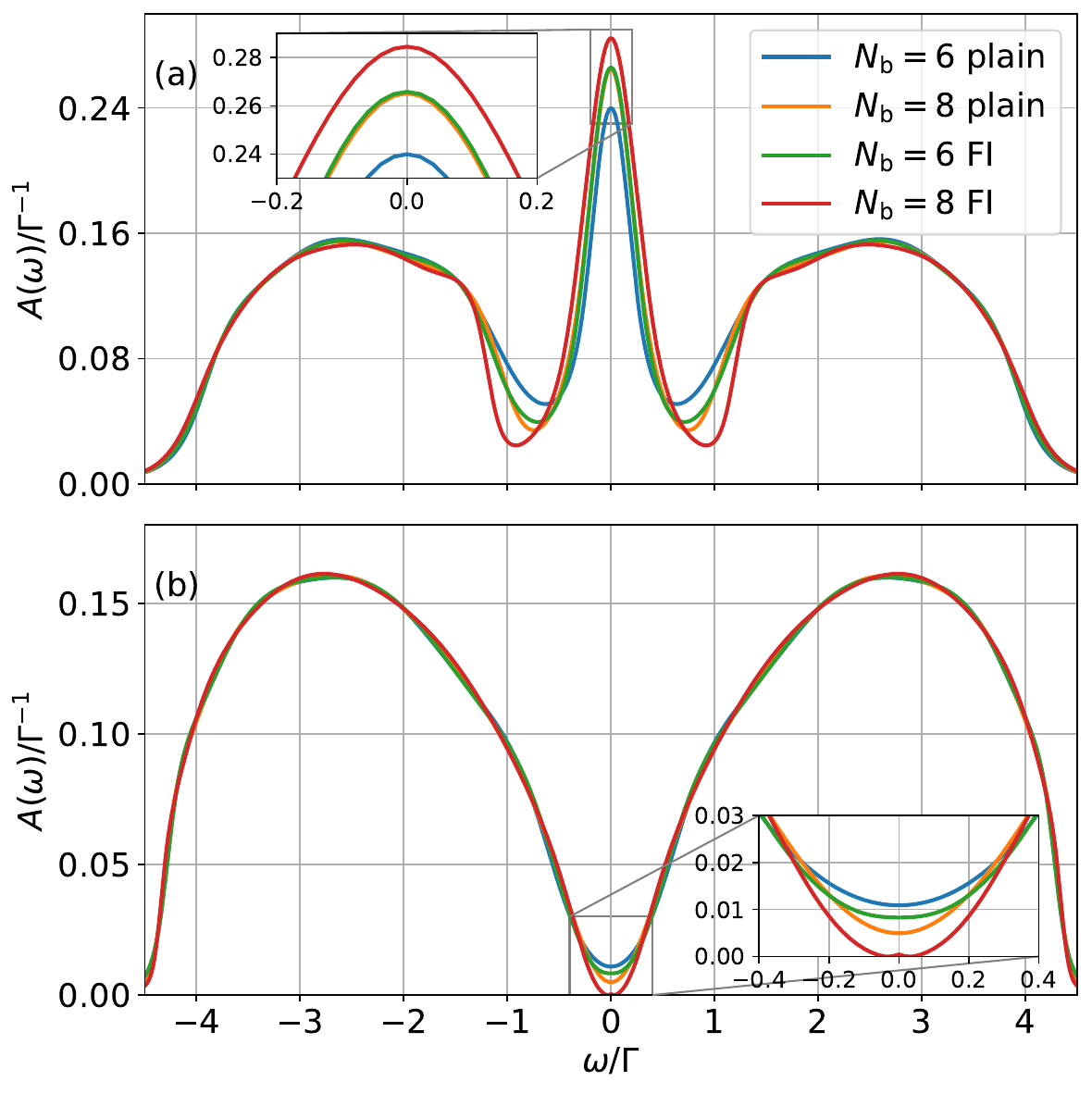}
	\caption{Spectral function for the infinite-dimensional Bethe lattice at $T/\Gamma = 0.05$, $\varepsilon_\text{imp} = -U/2$, $t/\Gamma = 1$, and (a) $U/\Gamma = 4.4$,
		i.e. in the metallic phase, or (b) for a point in the insulating phase at $U/\Gamma = 5$.}
	\label{fig:metal_ins}
\end{figure}

As mentioned above, the greatest advantage of the approach consists in its application as an efficient impurity solver for steady-state nonequilibrium DMFT. The method is also quite convenient as a real-frequency impurity solver in the equilibrium case.
To illustrate this, we apply it to the infinite-dimensional Hubbard model with the Bethe lattice DOS. We start
by addressing the most
challenging regime
in the equilibrium case, i.e. the vicinity of the metal-insulator transition.
We take a point in the metallic phase at
$T/\Gamma = 0.05$ and $U/\Gamma = 4.4$ and one nearby for
$U/\Gamma = 5.0$ and the same temperature in the insulating phase.
The corresponding spectral functions
are plotted in figure~\ref{fig:metal_ins}.
In both cases the results obtained for $N_\text{b} = 6$ FI lead to a significant improvement, i.e., they give a solution closer to $N_\text{b} = 8$ than the plain $N_\text{b} = 6$.
At the same time, the $N_\text{b} = 8$ FI result shows a more pronounced quasiparticle peak in the metallic and a clearer gap in the insulating phase.

\begin{figure}
	\centering
	\includegraphics[scale=0.47]{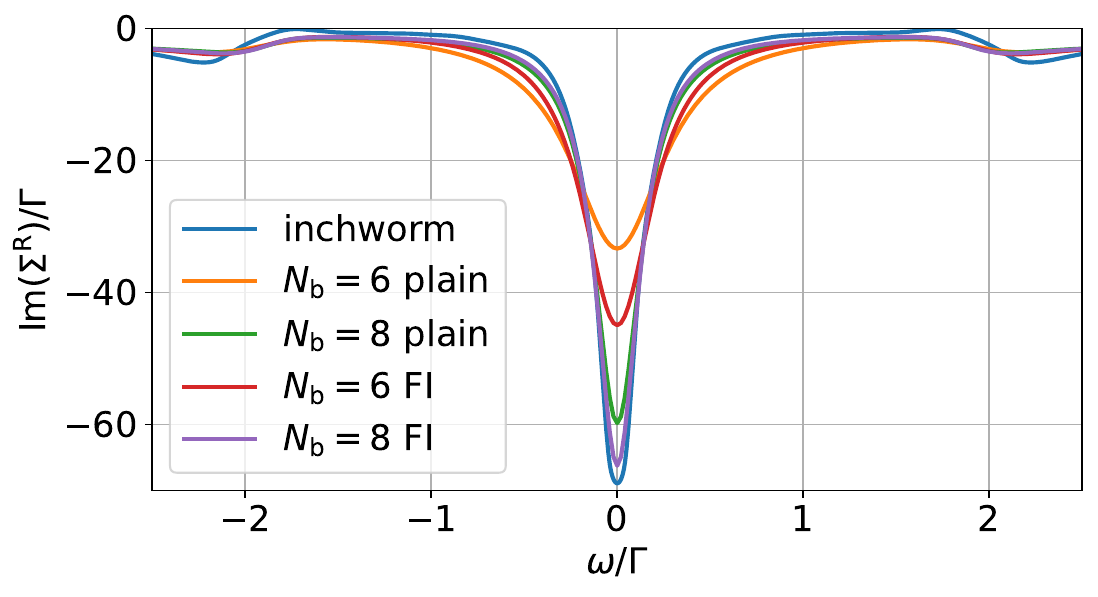}
	\caption{Imaginary part of the self-energy
		of a cold photodoped quasi steady state
		obtained by the FI approach and plain AMEA compared with inchworm QMC data (extracted from Ref.~\cite{ku.er.24}).
		Parameters are as Fig. 2b of Ref.~\cite{ku.er.24}.
	}
	\label{fig:inch_se}
\end{figure}

As a benchmark for a steady-state nonequilibrium situation we address cold photodoped states in a Mott insulator~\cite{da.li.21,ro.ra.08,se.pe.10,mu.ta.22}.
In Ref.~\cite{ku.er.24} it has been shown that these long-lived states can be reliably described by a
time-local quasi-steady-state ansatz for which numerically exact inchworm Monte Carlo data have been produced.
In Fig.~\ref{fig:inch_se} we compare the self-energy obtained by AMEA with $N_\text{b}=6$ and $8$ along with the corresponding FI corrections with the inchworm data from Ref.~\cite{ku.er.24}.
As one can see, the FI $N_\text{b}=8$ calculation produces a peak which is very close to the inchworm one.
On the other hand, the vanishing of the self-energy at the quasiparticle chemical potential around
$\omega/\Gamma \approx \pm 1.82$ is harder to reproduce. This is probably due to the fact that since the self-energy vanishes quadratically there, some kind of quadratic (instead of linear, as in \eqref{fi}) extrapolation should be carried out.

\paragraph{Conclusion}
We have introduced an approach for solving many-body impurity problems, particularly suited for strongly correlated systems out of equilibrium within dynamical mean-field theory. The method involves exactly solving the impurity embedded in a small auxiliary bath coupled to a Markovian environment, then applying a linear correction for the difference between the auxiliary and physical baths. This correction is well-justified when the difference between the two baths is small, as achieved with $N_\text{b}=6-8$. Our approach significantly improves results over a plain AMEA DMFT impurity solver in and out of equilibrium.

One key advantage of the present FI approach is that good accuracy is achieved with as few as $N_\text{b}=6$ bath sites, significantly speeding up impurity computations, which is crucial for DMFT self-consistency. Moreover, the ability to use fewer bath sites makes the method extendable to multi-orbital systems, which is essential for first-principles studies of correlated materials.

While the ``FI'' expression \eqref{sigmaimp} is straightforward, in practice the achieved accuracy sometimes depends on the choice of the batch of $N_\text{T}$ auxiliary hybridization functions. Our
extensive analysis for $N_\text{b}=4$ suggests that
a substantial improvement is obtained already with $N_\text{T}=2$.
The question is whether this conclusion is valid (a) also for larger $N_\text{b}$ and (b) for more structured hybridization functions. Increasing $N_\text{T}$ improves the accuracy of the self-energy in \eqref{sigmaimp} because it reduces $|\mmu|$. However, when $|\mmu|$ becomes smaller or of the order of $|\meps|^2$, the error gets dominated by the latter and there is no point of further reducing $|\mmu|$. This may suggest that larger $N_\text{T}$ could be necessary when using $N_\text{b} \gtrsim 6$ for which $|\meps|$ is smaller.
In the case of a structured (physical) hybrization function, the FI approach could provide a receipt to cleverly
combine the result of auxiliary hybridization functions adapted to one or to the other region. In this case, increasing $N_\text{T}$ should also improve the results. These issues should be addressed in future studies.

\paragraph{Acknowledgments}

This research was funded by the Austrian Science Fund (FWF) [Grant DOI:10.55776/P33165], and by NaWi Graz.

Results have been obtained using the A-Cluster at TU Graz
as well as the Vienna Scientific Cluster.

\bibliography{references_database, additional_citations}

\begin{thebibliography}{68}%
\makeatletter
\providecommand \@ifxundefined [1]{%
 \@ifx{#1\undefined}
}%
\providecommand \@ifnum [1]{%
 \ifnum #1\expandafter \@firstoftwo
 \else \expandafter \@secondoftwo
 \fi
}%
\providecommand \@ifx [1]{%
 \ifx #1\expandafter \@firstoftwo
 \else \expandafter \@secondoftwo
 \fi
}%
\providecommand \natexlab [1]{#1}%
\providecommand \enquote  [1]{``#1''}%
\providecommand \bibnamefont  [1]{#1}%
\providecommand \bibfnamefont [1]{#1}%
\providecommand \citenamefont [1]{#1}%
\providecommand \href@noop [0]{\@secondoftwo}%
\providecommand \href [0]{\begingroup \@sanitize@url \@href}%
\providecommand \@href[1]{\@@startlink{#1}\@@href}%
\providecommand \@@href[1]{\endgroup#1\@@endlink}%
\providecommand \@sanitize@url [0]{\catcode `\\12\catcode `\$12\catcode
  `\&12\catcode `\#12\catcode `\^12\catcode `\_12\catcode `\%12\relax}%
\providecommand \@@startlink[1]{}%
\providecommand \@@endlink[0]{}%
\providecommand \url  [0]{\begingroup\@sanitize@url \@url }%
\providecommand \@url [1]{\endgroup\@href {#1}{\urlprefix }}%
\providecommand \urlprefix  [0]{URL }%
\providecommand \Eprint [0]{\href }%
\providecommand \doibase [0]{https://doi.org/}%
\providecommand \selectlanguage [0]{\@gobble}%
\providecommand \bibinfo  [0]{\@secondoftwo}%
\providecommand \bibfield  [0]{\@secondoftwo}%
\providecommand \translation [1]{[#1]}%
\providecommand \BibitemOpen [0]{}%
\providecommand \bibitemStop [0]{}%
\providecommand \bibitemNoStop [0]{.\EOS\space}%
\providecommand \EOS [0]{\spacefactor3000\relax}%
\providecommand \BibitemShut  [1]{\csname bibitem#1\endcsname}%
\let\auto@bib@innerbib\@empty
\bibitem [{\citenamefont {Iwai}\ \emph {et~al.}(2003)\citenamefont {Iwai},
  \citenamefont {Ono}, \citenamefont {Maeda}, \citenamefont {Matsuzaki},
  \citenamefont {Kishida}, \citenamefont {Okamoto},\ and\ \citenamefont
  {Tokura}}]{iw.on.03}%
  \BibitemOpen
  \bibfield  {author} {\bibinfo {author} {\bibfnamefont {S.}~\bibnamefont
  {Iwai}}, \bibinfo {author} {\bibfnamefont {M.}~\bibnamefont {Ono}}, \bibinfo
  {author} {\bibfnamefont {A.}~\bibnamefont {Maeda}}, \bibinfo {author}
  {\bibfnamefont {H.}~\bibnamefont {Matsuzaki}}, \bibinfo {author}
  {\bibfnamefont {H.}~\bibnamefont {Kishida}}, \bibinfo {author} {\bibfnamefont
  {H.}~\bibnamefont {Okamoto}},\ and\ \bibinfo {author} {\bibfnamefont
  {Y.}~\bibnamefont {Tokura}},\ }\bibfield  {title} {\bibinfo {title}
  {Ultrafast optical switching to a metallic state by photoinduced mott
  transition in a halogen-bridged nickel-chain compound},\ }\href
  {https://doi.org/10.1103/PhysRevLett.91.057401} {\bibfield  {journal}
  {\bibinfo  {journal} {Phys. Rev. Lett.}\ }\textbf {\bibinfo {volume} {91}},\
  \bibinfo {pages} {057401} (\bibinfo {year} {2003})}\BibitemShut {NoStop}%
\bibitem [{\citenamefont {Cavalleri}\ \emph {et~al.}(2004)\citenamefont
  {Cavalleri}, \citenamefont {Dekorsy}, \citenamefont {Chong}, \citenamefont
  {Kieffer},\ and\ \citenamefont {Schoenlein}}]{ca.de.04}%
  \BibitemOpen
  \bibfield  {author} {\bibinfo {author} {\bibfnamefont {A.}~\bibnamefont
  {Cavalleri}}, \bibinfo {author} {\bibfnamefont {T.}~\bibnamefont {Dekorsy}},
  \bibinfo {author} {\bibfnamefont {H.~H.~W.}\ \bibnamefont {Chong}}, \bibinfo
  {author} {\bibfnamefont {J.~C.}\ \bibnamefont {Kieffer}},\ and\ \bibinfo
  {author} {\bibfnamefont {R.~W.}\ \bibnamefont {Schoenlein}},\ }\bibfield
  {title} {\bibinfo {title} {Evidence for a structurally-driven
  insulator-to-metal transition in {VO$_{2}$:} a view from the ultrafast
  timescale},\ }\href {https://doi.org/10.1103/PhysRevB.70.161102} {\bibfield
  {journal} {\bibinfo  {journal} {Phys. Rev. B}\ }\textbf {\bibinfo {volume}
  {70}},\ \bibinfo {pages} {161102} (\bibinfo {year} {2004})}\BibitemShut
  {NoStop}%
\bibitem [{\citenamefont {Perfetti}\ \emph {et~al.}(2006)\citenamefont
  {Perfetti}, \citenamefont {Loukakos}, \citenamefont {Lisowski}, \citenamefont
  {Bovensiepen}, \citenamefont {Berger}, \citenamefont {Biermann},
  \citenamefont {Cornaglia}, \citenamefont {Georges},\ and\ \citenamefont
  {Wolf}}]{pe.lo.06}%
  \BibitemOpen
  \bibfield  {author} {\bibinfo {author} {\bibfnamefont {L.}~\bibnamefont
  {Perfetti}}, \bibinfo {author} {\bibfnamefont {P.~A.}\ \bibnamefont
  {Loukakos}}, \bibinfo {author} {\bibfnamefont {M.}~\bibnamefont {Lisowski}},
  \bibinfo {author} {\bibfnamefont {U.}~\bibnamefont {Bovensiepen}}, \bibinfo
  {author} {\bibfnamefont {H.}~\bibnamefont {Berger}}, \bibinfo {author}
  {\bibfnamefont {S.}~\bibnamefont {Biermann}}, \bibinfo {author}
  {\bibfnamefont {P.~S.}\ \bibnamefont {Cornaglia}}, \bibinfo {author}
  {\bibfnamefont {A.}~\bibnamefont {Georges}},\ and\ \bibinfo {author}
  {\bibfnamefont {M.}~\bibnamefont {Wolf}},\ }\bibfield  {title} {\bibinfo
  {title} {Time evolution of the electronic structure of 1t-tas$_{2}$ through
  the insulator-metal transition},\ }\href
  {https://doi.org/10.1103/PhysRevLett.97.067402} {\bibfield  {journal}
  {\bibinfo  {journal} {Phys. Rev. Lett.}\ }\textbf {\bibinfo {volume} {97}},\
  \bibinfo {pages} {067402} (\bibinfo {year} {2006})}\BibitemShut {NoStop}%
\bibitem [{\citenamefont {Fausti}\ \emph {et~al.}(2011)\citenamefont {Fausti},
  \citenamefont {Tobey}, \citenamefont {Dean}, \citenamefont {Kaiser},
  \citenamefont {Dienst}, \citenamefont {Hoffmann}, \citenamefont {Pyon},
  \citenamefont {Takayama}, \citenamefont {Takagi},\ and\ \citenamefont
  {Cavalleri}}]{fa.to.11}%
  \BibitemOpen
  \bibfield  {author} {\bibinfo {author} {\bibfnamefont {D.}~\bibnamefont
  {Fausti}}, \bibinfo {author} {\bibfnamefont {R.~I.}\ \bibnamefont {Tobey}},
  \bibinfo {author} {\bibfnamefont {N.}~\bibnamefont {Dean}}, \bibinfo {author}
  {\bibfnamefont {S.}~\bibnamefont {Kaiser}}, \bibinfo {author} {\bibfnamefont
  {A.}~\bibnamefont {Dienst}}, \bibinfo {author} {\bibfnamefont {M.~C.}\
  \bibnamefont {Hoffmann}}, \bibinfo {author} {\bibfnamefont {S.}~\bibnamefont
  {Pyon}}, \bibinfo {author} {\bibfnamefont {T.}~\bibnamefont {Takayama}},
  \bibinfo {author} {\bibfnamefont {H.}~\bibnamefont {Takagi}},\ and\ \bibinfo
  {author} {\bibfnamefont {A.}~\bibnamefont {Cavalleri}},\ }\bibfield  {title}
  {\bibinfo {title} {{Light-Induced} superconductivity in a {Stripe-Ordered}
  cuprate},\ }\href {https://doi.org/10.1126/science.1197294} {\bibfield
  {journal} {\bibinfo  {journal} {Science}\ }\textbf {\bibinfo {volume}
  {331}},\ \bibinfo {pages} {189 } (\bibinfo {year} {2011})}\BibitemShut
  {NoStop}%
\bibitem [{\citenamefont {Zutic}\ \emph {et~al.}(2004)\citenamefont {Zutic},
  \citenamefont {Fabian},\ and\ \citenamefont {Sarma}}]{zu.fa.04}%
  \BibitemOpen
  \bibfield  {author} {\bibinfo {author} {\bibfnamefont {I.}~\bibnamefont
  {Zutic}}, \bibinfo {author} {\bibfnamefont {J.}~\bibnamefont {Fabian}},\ and\
  \bibinfo {author} {\bibfnamefont {S.~D.}\ \bibnamefont {Sarma}},\ }\bibfield
  {title} {\bibinfo {title} {Spintronics: Fundamentals and applications},\
  }\href {https://doi.org/10.1103/RevModPhys.76.323} {\bibfield  {journal}
  {\bibinfo  {journal} {Rev. Mod. Phys.}\ }\textbf {\bibinfo {volume} {76}},\
  \bibinfo {pages} {323} (\bibinfo {year} {2004})}\BibitemShut {NoStop}%
\bibitem [{\citenamefont {Bonilla}\ and\ \citenamefont
  {Grahn}(2005)}]{bo.gr.05}%
  \BibitemOpen
  \bibfield  {author} {\bibinfo {author} {\bibfnamefont {L.~L.}\ \bibnamefont
  {Bonilla}}\ and\ \bibinfo {author} {\bibfnamefont {H.~T.}\ \bibnamefont
  {Grahn}},\ }\bibfield  {title} {\bibinfo {title} {Non-linear dynamics of
  semiconductor superlattices},\ }\href
  {http://stacks.iop.org/0034-4885/68/i=3/a=R03} {\bibfield  {journal}
  {\bibinfo  {journal} {Rep. Prog. Phys.}\ }\textbf {\bibinfo {volume} {68}},\
  \bibinfo {pages} {577} (\bibinfo {year} {2005})}\BibitemShut {NoStop}%
\bibitem [{\citenamefont {Raizen}\ \emph {et~al.}(1997)\citenamefont {Raizen},
  \citenamefont {Salomon},\ and\ \citenamefont {Niu}}]{ra.sa.97}%
  \BibitemOpen
  \bibfield  {author} {\bibinfo {author} {\bibfnamefont {M.}~\bibnamefont
  {Raizen}}, \bibinfo {author} {\bibfnamefont {C.}~\bibnamefont {Salomon}},\
  and\ \bibinfo {author} {\bibfnamefont {Q.}~\bibnamefont {Niu}},\ }\bibfield
  {title} {\bibinfo {title} {New light on quantum transport},\ }\href
  {https://doi.org/10.1063/1.881845} {\bibfield  {journal} {\bibinfo  {journal}
  {Phys. Today}\ }\textbf {\bibinfo {volume} {50}},\ \bibinfo {pages} {30}
  (\bibinfo {year} {1997})}\BibitemShut {NoStop}%
\bibitem [{\citenamefont {Jaksch}\ \emph {et~al.}(1998)\citenamefont {Jaksch},
  \citenamefont {Bruder}, \citenamefont {Cirac}, \citenamefont {Gardiner},\
  and\ \citenamefont {Zoller}}]{ja.br.98}%
  \BibitemOpen
  \bibfield  {author} {\bibinfo {author} {\bibfnamefont {D.}~\bibnamefont
  {Jaksch}}, \bibinfo {author} {\bibfnamefont {C.}~\bibnamefont {Bruder}},
  \bibinfo {author} {\bibfnamefont {J.~I.}\ \bibnamefont {Cirac}}, \bibinfo
  {author} {\bibfnamefont {C.~W.}\ \bibnamefont {Gardiner}},\ and\ \bibinfo
  {author} {\bibfnamefont {P.}~\bibnamefont {Zoller}},\ }\bibfield  {title}
  {\bibinfo {title} {Cold bosonic atoms in optical lattices},\ }\href@noop {}
  {\bibfield  {journal} {\bibinfo  {journal} {Phys. Rev. Lett.}\ }\textbf
  {\bibinfo {volume} {81}},\ \bibinfo {pages} {3108} (\bibinfo {year}
  {1998})}\BibitemShut {NoStop}%
\bibitem [{\citenamefont {Greiner}\ \emph {et~al.}(2002)\citenamefont
  {Greiner}, \citenamefont {Mandel}, \citenamefont {Esslinger}, \citenamefont
  {H{\"a}nsch},\ and\ \citenamefont {Bloch}}]{gr.ma.02}%
  \BibitemOpen
  \bibfield  {author} {\bibinfo {author} {\bibfnamefont {M.}~\bibnamefont
  {Greiner}}, \bibinfo {author} {\bibfnamefont {O.}~\bibnamefont {Mandel}},
  \bibinfo {author} {\bibfnamefont {T.}~\bibnamefont {Esslinger}}, \bibinfo
  {author} {\bibfnamefont {T.~W.}\ \bibnamefont {H{\"a}nsch}},\ and\ \bibinfo
  {author} {\bibfnamefont {I.}~\bibnamefont {Bloch}},\ }\bibfield  {title}
  {\bibinfo {title} {Quantum phase transition from a superfluid to a mott
  insulator in a gas of ultracold atoms},\ }\href@noop {} {\bibfield  {journal}
  {\bibinfo  {journal} {Nature}\ }\textbf {\bibinfo {volume} {415}},\ \bibinfo
  {pages} {39} (\bibinfo {year} {2002})}\BibitemShut {NoStop}%
\bibitem [{\citenamefont {Hartmann}\ \emph {et~al.}(2008)\citenamefont
  {Hartmann}, \citenamefont {Brand{\~a}o},\ and\ \citenamefont
  {Plenio}}]{ha.br.08}%
  \BibitemOpen
  \bibfield  {author} {\bibinfo {author} {\bibfnamefont {M.}~\bibnamefont
  {Hartmann}}, \bibinfo {author} {\bibfnamefont {F.}~\bibnamefont
  {Brand{\~a}o}},\ and\ \bibinfo {author} {\bibfnamefont {M.}~\bibnamefont
  {Plenio}},\ }\bibfield  {title} {\bibinfo {title} {Quantum many-body
  phenomena in coupled cavity arrays},\ }\href
  {https://doi.org/10.1002/lpor.200810046} {\bibfield  {journal} {\bibinfo
  {journal} {Laser {\&} Photonics Rev.}\ }\textbf {\bibinfo {volume} {2}},\
  \bibinfo {pages} {527} (\bibinfo {year} {2008})}\BibitemShut {NoStop}%
\bibitem [{\citenamefont {Mitra}\ \emph {et~al.}(2006)\citenamefont {Mitra},
  \citenamefont {Takei}, \citenamefont {Kim},\ and\ \citenamefont
  {Millis}}]{mi.ta.06}%
  \BibitemOpen
  \bibfield  {author} {\bibinfo {author} {\bibfnamefont {A.}~\bibnamefont
  {Mitra}}, \bibinfo {author} {\bibfnamefont {S.}~\bibnamefont {Takei}},
  \bibinfo {author} {\bibfnamefont {Y.~B.}\ \bibnamefont {Kim}},\ and\ \bibinfo
  {author} {\bibfnamefont {A.~J.}\ \bibnamefont {Millis}},\ }\bibfield  {title}
  {\bibinfo {title} {Nonequilibrium quantum criticality in open electronic
  systems},\ }\href@noop {} {\bibfield  {journal} {\bibinfo  {journal} {Phys.
  Rev. Lett.}\ }\textbf {\bibinfo {volume} {97}},\ \bibinfo {pages} {236808}
  (\bibinfo {year} {2006})}\BibitemShut {NoStop}%
\bibitem [{\citenamefont {Cazalilla}(2006)}]{caza.06}%
  \BibitemOpen
  \bibfield  {author} {\bibinfo {author} {\bibfnamefont {M.~A.}\ \bibnamefont
  {Cazalilla}},\ }\bibfield  {title} {\bibinfo {title} {Effect of suddenly
  turning on interactions in the luttinger model},\ }\href
  {https://doi.org/10.1103/PhysRevLett.97.156403} {\bibfield  {journal}
  {\bibinfo  {journal} {Phys. Rev. Lett.}\ }\textbf {\bibinfo {volume} {97}},\
  \bibinfo {pages} {156403} (\bibinfo {year} {2006})}\BibitemShut {NoStop}%
\bibitem [{\citenamefont {Calabrese}\ and\ \citenamefont
  {Cardy}(2007)}]{ca.ca.07}%
  \BibitemOpen
  \bibfield  {author} {\bibinfo {author} {\bibfnamefont {P.}~\bibnamefont
  {Calabrese}}\ and\ \bibinfo {author} {\bibfnamefont {J.}~\bibnamefont
  {Cardy}},\ }\bibfield  {title} {\bibinfo {title} {Quantum quenches in
  extended systems},\ }\href@noop {} {\bibfield  {journal} {\bibinfo  {journal}
  {J. Stat. Mech.}\ }\textbf {\bibinfo {volume} {2007}},\ \bibinfo {pages}
  {P06008} (\bibinfo {year} {2007})}\BibitemShut {NoStop}%
\bibitem [{\citenamefont {Rigol}\ \emph {et~al.}(2008)\citenamefont {Rigol},
  \citenamefont {Dunjko},\ and\ \citenamefont {Olshanii}}]{ri.du.08}%
  \BibitemOpen
  \bibfield  {author} {\bibinfo {author} {\bibfnamefont {M.}~\bibnamefont
  {Rigol}}, \bibinfo {author} {\bibfnamefont {V.}~\bibnamefont {Dunjko}},\ and\
  \bibinfo {author} {\bibfnamefont {M.}~\bibnamefont {Olshanii}},\ }\bibfield
  {title} {\bibinfo {title} {Thermalization and its mechanism for generic
  isolated quantum systems},\ }\href {https://doi.org/10.1038/nature06838}
  {\bibfield  {journal} {\bibinfo  {journal} {Nature}\ }\textbf {\bibinfo
  {volume} {452}},\ \bibinfo {pages} {854} (\bibinfo {year}
  {2008})}\BibitemShut {NoStop}%
\bibitem [{\citenamefont {Leggett}\ \emph {et~al.}(1987)\citenamefont
  {Leggett}, \citenamefont {Chakravarty}, \citenamefont {Dorsey}, \citenamefont
  {Fisher}, \citenamefont {Garg},\ and\ \citenamefont {Zwerger}}]{le.ch.87}%
  \BibitemOpen
  \bibfield  {author} {\bibinfo {author} {\bibfnamefont {A.~J.}\ \bibnamefont
  {Leggett}}, \bibinfo {author} {\bibfnamefont {S.}~\bibnamefont
  {Chakravarty}}, \bibinfo {author} {\bibfnamefont {A.~T.}\ \bibnamefont
  {Dorsey}}, \bibinfo {author} {\bibfnamefont {M.~P.~A.}\ \bibnamefont
  {Fisher}}, \bibinfo {author} {\bibfnamefont {A.}~\bibnamefont {Garg}},\ and\
  \bibinfo {author} {\bibfnamefont {W.}~\bibnamefont {Zwerger}},\ }\bibfield
  {title} {\bibinfo {title} {Dynamics of the dissipative two-state system},\
  }\href@noop {} {\bibfield  {journal} {\bibinfo  {journal} {Rev. Mod. Phys.}\
  }\textbf {\bibinfo {volume} {59}},\ \bibinfo {pages} {1} (\bibinfo {year}
  {1987})}\BibitemShut {NoStop}%
\bibitem [{\citenamefont {Georges}\ \emph {et~al.}(1996)\citenamefont
  {Georges}, \citenamefont {Kotliar}, \citenamefont {Krauth},\ and\
  \citenamefont {Rozenberg}}]{ge.ko.96}%
  \BibitemOpen
  \bibfield  {author} {\bibinfo {author} {\bibfnamefont {A.}~\bibnamefont
  {Georges}}, \bibinfo {author} {\bibfnamefont {G.}~\bibnamefont {Kotliar}},
  \bibinfo {author} {\bibfnamefont {W.}~\bibnamefont {Krauth}},\ and\ \bibinfo
  {author} {\bibfnamefont {M.~J.}\ \bibnamefont {Rozenberg}},\ }\bibfield
  {title} {\bibinfo {title} {Dynamical mean-field theory of strongly correlated
  fermion systems and the limit of infinite dimensions},\ }\href
  {https://doi.org/10.1103/RevModPhys.68.13} {\bibfield  {journal} {\bibinfo
  {journal} {Rev. Mod. Phys.}\ }\textbf {\bibinfo {volume} {68}},\ \bibinfo
  {pages} {13} (\bibinfo {year} {1996})}\BibitemShut {NoStop}%
\bibitem [{\citenamefont {Metzner}\ and\ \citenamefont
  {Vollhardt}(1989)}]{me.vo.89}%
  \BibitemOpen
  \bibfield  {author} {\bibinfo {author} {\bibfnamefont {W.}~\bibnamefont
  {Metzner}}\ and\ \bibinfo {author} {\bibfnamefont {D.}~\bibnamefont
  {Vollhardt}},\ }\bibfield  {title} {\bibinfo {title} {Correlated lattice
  fermions in $d=\infty$ dimensions},\ }\href
  {https://doi.org/10.1103/PhysRevLett.62.324} {\bibfield  {journal} {\bibinfo
  {journal} {Phys. Rev. Lett.}\ }\textbf {\bibinfo {volume} {62}},\ \bibinfo
  {pages} {324} (\bibinfo {year} {1989})}\BibitemShut {NoStop}%
\bibitem [{\citenamefont {Schmidt}\ and\ \citenamefont
  {Monien}(2002)}]{sc.mo.02u}%
  \BibitemOpen
  \bibfield  {author} {\bibinfo {author} {\bibfnamefont {P.}~\bibnamefont
  {Schmidt}}\ and\ \bibinfo {author} {\bibfnamefont {H.}~\bibnamefont
  {Monien}},\ }\bibfield  {title} {\bibinfo {title} {Nonequilibrium dynamical
  mean -- field theory of a strongly correlated system}} (\bibinfo {year}
  {2002}),\ \bibinfo {note} {cond-mat/0202046}\BibitemShut {NoStop}%
\bibitem [{\citenamefont {Freericks}\ \emph {et~al.}(2006)\citenamefont
  {Freericks}, \citenamefont {Turkowski},\ and\ \citenamefont
  {Zlati{\'{c}}}}]{fr.tu.06}%
  \BibitemOpen
  \bibfield  {author} {\bibinfo {author} {\bibfnamefont {J.~K.}\ \bibnamefont
  {Freericks}}, \bibinfo {author} {\bibfnamefont {V.~M.}\ \bibnamefont
  {Turkowski}},\ and\ \bibinfo {author} {\bibfnamefont {V.}~\bibnamefont
  {Zlati{\'{c}}}},\ }\bibfield  {title} {\bibinfo {title} {Nonequilibrium
  dynamical mean-field theory},\ }\href
  {https://doi.org/10.1103/PhysRevLett.97.266408} {\bibfield  {journal}
  {\bibinfo  {journal} {Phys. Rev. Lett.}\ }\textbf {\bibinfo {volume} {97}},\
  \bibinfo {pages} {266408} (\bibinfo {year} {2006})}\BibitemShut {NoStop}%
\bibitem [{\citenamefont {Freericks}(2008)}]{free.08}%
  \BibitemOpen
  \bibfield  {author} {\bibinfo {author} {\bibfnamefont {J.~K.}\ \bibnamefont
  {Freericks}},\ }\bibfield  {title} {\bibinfo {title} {Quenching bloch
  oscillations in a strongly correlated material: Nonequilibrium dynamical
  mean-field theory},\ }\href {https://doi.org/10.1103/PhysRevB.77.075109}
  {\bibfield  {journal} {\bibinfo  {journal} {Phys. Rev. B}\ }\textbf {\bibinfo
  {volume} {77}},\ \bibinfo {pages} {075109} (\bibinfo {year}
  {2008})}\BibitemShut {NoStop}%
\bibitem [{\citenamefont {Joura}\ \emph {et~al.}(2008)\citenamefont {Joura},
  \citenamefont {Freericks},\ and\ \citenamefont {Pruschke}}]{jo.fr.08}%
  \BibitemOpen
  \bibfield  {author} {\bibinfo {author} {\bibfnamefont {A.~V.}\ \bibnamefont
  {Joura}}, \bibinfo {author} {\bibfnamefont {J.~K.}\ \bibnamefont
  {Freericks}},\ and\ \bibinfo {author} {\bibfnamefont {T.}~\bibnamefont
  {Pruschke}},\ }\bibfield  {title} {\bibinfo {title} {Steady-state
  nonequilibrium density of states of driven strongly correlated lattice models
  in infinite dimensions},\ }\href
  {https://doi.org/10.1103/PhysRevLett.101.196401} {\bibfield  {journal}
  {\bibinfo  {journal} {Phys. Rev. Lett.}\ }\textbf {\bibinfo {volume} {101}},\
  \bibinfo {pages} {196401} (\bibinfo {year} {2008})}\BibitemShut {NoStop}%
\bibitem [{\citenamefont {Eckstein}\ \emph {et~al.}(2009)\citenamefont
  {Eckstein}, \citenamefont {Kollar},\ and\ \citenamefont {Werner}}]{ec.ko.09}%
  \BibitemOpen
  \bibfield  {author} {\bibinfo {author} {\bibfnamefont {M.}~\bibnamefont
  {Eckstein}}, \bibinfo {author} {\bibfnamefont {M.}~\bibnamefont {Kollar}},\
  and\ \bibinfo {author} {\bibfnamefont {P.}~\bibnamefont {Werner}},\
  }\bibfield  {title} {\bibinfo {title} {Thermalization after an interaction
  quench in the hubbard model},\ }\href
  {https://doi.org/10.1103/PhysRevLett.103.056403} {\bibfield  {journal}
  {\bibinfo  {journal} {Phys. Rev. Lett.}\ }\textbf {\bibinfo {volume} {103}},\
  \bibinfo {pages} {056403} (\bibinfo {year} {2009})}\BibitemShut {NoStop}%
\bibitem [{\citenamefont {Okamoto}(2007)}]{okam.07}%
  \BibitemOpen
  \bibfield  {author} {\bibinfo {author} {\bibfnamefont {S.}~\bibnamefont
  {Okamoto}},\ }\bibfield  {title} {\bibinfo {title} {Nonequilibrium transport
  and optical properties of model metal-mott-insulator-metal
  heterostructures},\ }\href {https://doi.org/10.1103/PhysRevB.76.035105}
  {\bibfield  {journal} {\bibinfo  {journal} {Phys. Rev. B}\ }\textbf {\bibinfo
  {volume} {76}},\ \bibinfo {pages} {035105} (\bibinfo {year}
  {2007})}\BibitemShut {NoStop}%
\bibitem [{\citenamefont {Aoki}\ \emph {et~al.}(2014)\citenamefont {Aoki},
  \citenamefont {Tsuji}, \citenamefont {Eckstein}, \citenamefont {Kollar},
  \citenamefont {Oka},\ and\ \citenamefont {Werner}}]{ao.ts.14}%
  \BibitemOpen
  \bibfield  {author} {\bibinfo {author} {\bibfnamefont {H.}~\bibnamefont
  {Aoki}}, \bibinfo {author} {\bibfnamefont {N.}~\bibnamefont {Tsuji}},
  \bibinfo {author} {\bibfnamefont {M.}~\bibnamefont {Eckstein}}, \bibinfo
  {author} {\bibfnamefont {M.}~\bibnamefont {Kollar}}, \bibinfo {author}
  {\bibfnamefont {T.}~\bibnamefont {Oka}},\ and\ \bibinfo {author}
  {\bibfnamefont {P.}~\bibnamefont {Werner}},\ }\bibfield  {title} {\bibinfo
  {title} {Nonequilibrium dynamical mean-field theory and its applications},\
  }\href {https://doi.org/10.1103/RevModPhys.86.779} {\bibfield  {journal}
  {\bibinfo  {journal} {Rev. Mod. Phys.}\ }\textbf {\bibinfo {volume} {86}},\
  \bibinfo {pages} {779} (\bibinfo {year} {2014})}\BibitemShut {NoStop}%
\bibitem [{\citenamefont {Gull}\ \emph {et~al.}(2011)\citenamefont {Gull},
  \citenamefont {Millis}, \citenamefont {Lichtenstein}, \citenamefont
  {Rubtsov}, \citenamefont {Troyer},\ and\ \citenamefont {Werner}}]{gu.mi.11}%
  \BibitemOpen
  \bibfield  {author} {\bibinfo {author} {\bibfnamefont {E.}~\bibnamefont
  {Gull}}, \bibinfo {author} {\bibfnamefont {A.~J.}\ \bibnamefont {Millis}},
  \bibinfo {author} {\bibfnamefont {A.~I.}\ \bibnamefont {Lichtenstein}},
  \bibinfo {author} {\bibfnamefont {A.~N.}\ \bibnamefont {Rubtsov}}, \bibinfo
  {author} {\bibfnamefont {M.}~\bibnamefont {Troyer}},\ and\ \bibinfo {author}
  {\bibfnamefont {P.}~\bibnamefont {Werner}},\ }\bibfield  {title} {\bibinfo
  {title} {Continuous-time monte carlo methods for quantum impurity models},\
  }\href {https://doi.org/10.1103/RevModPhys.83.349} {\bibfield  {journal}
  {\bibinfo  {journal} {Rev. Mod. Phys.}\ }\textbf {\bibinfo {volume} {83}},\
  \bibinfo {pages} {349} (\bibinfo {year} {2011})}\BibitemShut {NoStop}%
\bibitem [{\citenamefont {Werner}\ \emph {et~al.}(2009)\citenamefont {Werner},
  \citenamefont {Oka},\ and\ \citenamefont {Millis}}]{we.ok.09}%
  \BibitemOpen
  \bibfield  {author} {\bibinfo {author} {\bibfnamefont {P.}~\bibnamefont
  {Werner}}, \bibinfo {author} {\bibfnamefont {T.}~\bibnamefont {Oka}},\ and\
  \bibinfo {author} {\bibfnamefont {A.~J.}\ \bibnamefont {Millis}},\ }\bibfield
   {title} {\bibinfo {title} {Diagrammatic monte carlo simulation of
  nonequilibrium systems},\ }\href {https://doi.org/10.1103/PhysRevB.79.035320}
  {\bibfield  {journal} {\bibinfo  {journal} {Phys. Rev. B}\ }\textbf {\bibinfo
  {volume} {79}},\ \bibinfo {pages} {035320} (\bibinfo {year}
  {2009})}\BibitemShut {NoStop}%
\bibitem [{\citenamefont {White}\ and\ \citenamefont
  {Feiguin}(2004)}]{wh.fe.04}%
  \BibitemOpen
  \bibfield  {author} {\bibinfo {author} {\bibfnamefont {S.~R.}\ \bibnamefont
  {White}}\ and\ \bibinfo {author} {\bibfnamefont {A.~E.}\ \bibnamefont
  {Feiguin}},\ }\bibfield  {title} {\bibinfo {title} {Real-time evolution using
  the density matrix renormalization group},\ }\href
  {https://doi.org/10.1103/PhysRevLett.93.076401} {\bibfield  {journal}
  {\bibinfo  {journal} {Phys. Rev. Lett.}\ }\textbf {\bibinfo {volume} {93}},\
  \bibinfo {pages} {076401} (\bibinfo {year} {2004})}\BibitemShut {NoStop}%
\bibitem [{\citenamefont {Daley}\ \emph {et~al.}(2004)\citenamefont {Daley},
  \citenamefont {Kollath}, \citenamefont {Schollw{\"o}ck},\ and\ \citenamefont
  {Vidal}}]{da.ko.04}%
  \BibitemOpen
  \bibfield  {author} {\bibinfo {author} {\bibfnamefont {A.~J.}\ \bibnamefont
  {Daley}}, \bibinfo {author} {\bibfnamefont {C.}~\bibnamefont {Kollath}},
  \bibinfo {author} {\bibfnamefont {U.}~\bibnamefont {Schollw{\"o}ck}},\ and\
  \bibinfo {author} {\bibfnamefont {G.}~\bibnamefont {Vidal}},\ }\bibfield
  {title} {\bibinfo {title} {Time-dependent density-matrix
  renormalization-group using adaptive effective hilbert spaces},\ }\href
  {http://stacks.iop.org/1742-5468/2004/i=04/a=P04005} {\bibfield  {journal}
  {\bibinfo  {journal} {J. Stat. Mech.}\ }\textbf {\bibinfo {volume} {2004}},\
  \bibinfo {pages} {P04005} (\bibinfo {year} {2004})}\BibitemShut {NoStop}%
\bibitem [{\citenamefont {Erpenbeck}\ \emph {et~al.}(2023)\citenamefont
  {Erpenbeck}, \citenamefont {Gull},\ and\ \citenamefont {Cohen}}]{er.gu.23}%
  \BibitemOpen
  \bibfield  {author} {\bibinfo {author} {\bibfnamefont {A.}~\bibnamefont
  {Erpenbeck}}, \bibinfo {author} {\bibfnamefont {E.}~\bibnamefont {Gull}},\
  and\ \bibinfo {author} {\bibfnamefont {G.}~\bibnamefont {Cohen}},\ }\bibfield
   {title} {\bibinfo {title} {Quantum monte carlo method in the steady state},\
  }\href {https://doi.org/10.1103/PhysRevLett.130.186301} {\bibfield  {journal}
  {\bibinfo  {journal} {Phys. Rev. Lett.}\ }\textbf {\bibinfo {volume} {130}},\
  \bibinfo {pages} {186301} (\bibinfo {year} {2023})}\BibitemShut {NoStop}%
\bibitem [{\citenamefont {Mehta}\ and\ \citenamefont
  {Andrei}(2006)}]{me.an.06}%
  \BibitemOpen
  \bibfield  {author} {\bibinfo {author} {\bibfnamefont {P.}~\bibnamefont
  {Mehta}}\ and\ \bibinfo {author} {\bibfnamefont {N.}~\bibnamefont {Andrei}},\
  }\bibfield  {title} {\bibinfo {title} {Nonequilibrium transport in quantum
  impurity models: The bethe ansatz for open systems},\ }\href
  {https://doi.org/10.1103/PhysRevLett.96.216802} {\bibfield  {journal}
  {\bibinfo  {journal} {Phys. Rev. Lett.}\ }\textbf {\bibinfo {volume} {96}},\
  \bibinfo {pages} {216802} (\bibinfo {year} {2006})}\BibitemShut {NoStop}%
\bibitem [{\citenamefont {Anders}(2008)}]{ande.08}%
  \BibitemOpen
  \bibfield  {author} {\bibinfo {author} {\bibfnamefont {F.~B.}\ \bibnamefont
  {Anders}},\ }\bibfield  {title} {\bibinfo {title} {Steady-state currents
  through nanodevices: A scattering-states numerical renormalization-group
  approach to open quantum systems},\ }\href
  {https://doi.org/10.1103/PhysRevLett.101.066804} {\bibfield  {journal}
  {\bibinfo  {journal} {Phys. Rev. Lett.}\ }\textbf {\bibinfo {volume} {101}},\
  \bibinfo {pages} {066804} (\bibinfo {year} {2008})}\BibitemShut {NoStop}%
\bibitem [{\citenamefont {Arrigoni}\ \emph {et~al.}(2013)\citenamefont
  {Arrigoni}, \citenamefont {Knap},\ and\ \citenamefont {von~der
  Linden}}]{ar.kn.13}%
  \BibitemOpen
  \bibfield  {author} {\bibinfo {author} {\bibfnamefont {E.}~\bibnamefont
  {Arrigoni}}, \bibinfo {author} {\bibfnamefont {M.}~\bibnamefont {Knap}},\
  and\ \bibinfo {author} {\bibfnamefont {W.}~\bibnamefont {von~der Linden}},\
  }\bibfield  {title} {\bibinfo {title} {Nonequilibrium dynamical mean field
  theory: an auxiliary quantum master equation approach},\ }\href
  {https://doi.org/10.1103/PhysRevLett.110.086403} {\bibfield  {journal}
  {\bibinfo  {journal} {Phys. Rev. Lett.}\ }\textbf {\bibinfo {volume} {110}},\
  \bibinfo {pages} {086403} (\bibinfo {year} {2013})}\BibitemShut {NoStop}%
\bibitem [{\citenamefont {Dorda}\ \emph {et~al.}(2014)\citenamefont {Dorda},
  \citenamefont {Nuss}, \citenamefont {von~der Linden},\ and\ \citenamefont
  {Arrigoni}}]{do.nu.14}%
  \BibitemOpen
  \bibfield  {author} {\bibinfo {author} {\bibfnamefont {A.}~\bibnamefont
  {Dorda}}, \bibinfo {author} {\bibfnamefont {M.}~\bibnamefont {Nuss}},
  \bibinfo {author} {\bibfnamefont {W.}~\bibnamefont {von~der Linden}},\ and\
  \bibinfo {author} {\bibfnamefont {E.}~\bibnamefont {Arrigoni}},\ }\bibfield
  {title} {\bibinfo {title} {Auxiliary master equation approach to non --
  equilibrium correlated impurities},\ }\href
  {https://doi.org/10.1103/PhysRevB.89.165105} {\bibfield  {journal} {\bibinfo
  {journal} {Phys. Rev. B}\ }\textbf {\bibinfo {volume} {89}},\ \bibinfo
  {pages} {165105} (\bibinfo {year} {2014})}\BibitemShut {NoStop}%
\bibitem [{\citenamefont {Dasari}\ \emph {et~al.}(2021)\citenamefont {Dasari},
  \citenamefont {Li}, \citenamefont {Werner},\ and\ \citenamefont
  {Eckstein}}]{da.li.21}%
  \BibitemOpen
  \bibfield  {author} {\bibinfo {author} {\bibfnamefont {N.}~\bibnamefont
  {Dasari}}, \bibinfo {author} {\bibfnamefont {J.}~\bibnamefont {Li}}, \bibinfo
  {author} {\bibfnamefont {P.}~\bibnamefont {Werner}},\ and\ \bibinfo {author}
  {\bibfnamefont {M.}~\bibnamefont {Eckstein}},\ }\bibfield  {title} {\bibinfo
  {title} {Photoinduced strange metal with electron and hole quasiparticles},\
  }\href {https://doi.org/10.1103/PhysRevB.103.L201116} {\bibfield  {journal}
  {\bibinfo  {journal} {Phys. Rev. B}\ }\textbf {\bibinfo {volume} {103}},\
  \bibinfo {pages} {L201116} (\bibinfo {year} {2021})}\BibitemShut {NoStop}%
\bibitem [{\citenamefont {Rosch}\ \emph {et~al.}(2008)\citenamefont {Rosch},
  \citenamefont {Rasch}, \citenamefont {Binz},\ and\ \citenamefont
  {Vojta}}]{ro.ra.08}%
  \BibitemOpen
  \bibfield  {author} {\bibinfo {author} {\bibfnamefont {A.}~\bibnamefont
  {Rosch}}, \bibinfo {author} {\bibfnamefont {D.}~\bibnamefont {Rasch}},
  \bibinfo {author} {\bibfnamefont {B.}~\bibnamefont {Binz}},\ and\ \bibinfo
  {author} {\bibfnamefont {M.}~\bibnamefont {Vojta}},\ }\bibfield  {title}
  {\bibinfo {title} {Metastable superfluidity of repulsive fermionic atoms in
  optical lattices},\ }\href {https://doi.org/10.1103/PhysRevLett.101.265301}
  {\bibfield  {journal} {\bibinfo  {journal} {Phys. Rev. Lett.}\ }\textbf
  {\bibinfo {volume} {101}},\ \bibinfo {pages} {265301} (\bibinfo {year}
  {2008})}\BibitemShut {NoStop}%
\bibitem [{\citenamefont {Sensarma}\ \emph {et~al.}(2010)\citenamefont
  {Sensarma}, \citenamefont {Pekker}, \citenamefont {Altman}, \citenamefont
  {Demler}, \citenamefont {Strohmaier}, \citenamefont {Greif}, \citenamefont
  {J\"ordens}, \citenamefont {Tarruell}, \citenamefont {Moritz},\ and\
  \citenamefont {Esslinger}}]{se.pe.10}%
  \BibitemOpen
  \bibfield  {author} {\bibinfo {author} {\bibfnamefont {R.}~\bibnamefont
  {Sensarma}}, \bibinfo {author} {\bibfnamefont {D.}~\bibnamefont {Pekker}},
  \bibinfo {author} {\bibfnamefont {E.}~\bibnamefont {Altman}}, \bibinfo
  {author} {\bibfnamefont {E.}~\bibnamefont {Demler}}, \bibinfo {author}
  {\bibfnamefont {N.}~\bibnamefont {Strohmaier}}, \bibinfo {author}
  {\bibfnamefont {D.}~\bibnamefont {Greif}}, \bibinfo {author} {\bibfnamefont
  {R.}~\bibnamefont {J\"ordens}}, \bibinfo {author} {\bibfnamefont
  {L.}~\bibnamefont {Tarruell}}, \bibinfo {author} {\bibfnamefont
  {H.}~\bibnamefont {Moritz}},\ and\ \bibinfo {author} {\bibfnamefont
  {T.}~\bibnamefont {Esslinger}},\ }\bibfield  {title} {\bibinfo {title}
  {Lifetime of double occupancies in the fermi-hubbard model},\ }\href
  {https://doi.org/10.1103/PhysRevB.82.224302} {\bibfield  {journal} {\bibinfo
  {journal} {Phys. Rev. B}\ }\textbf {\bibinfo {volume} {82}},\ \bibinfo
  {pages} {224302} (\bibinfo {year} {2010})}\BibitemShut {NoStop}%
\bibitem [{\citenamefont {Murakami}\ \emph {et~al.}(2022)\citenamefont
  {Murakami}, \citenamefont {Takayoshi}, \citenamefont {Kaneko}, \citenamefont
  {Sun}, \citenamefont {Golez}, \citenamefont {Millis},\ and\ \citenamefont
  {Werner}}]{mu.ta.22}%
  \BibitemOpen
  \bibfield  {author} {\bibinfo {author} {\bibfnamefont {Y.}~\bibnamefont
  {Murakami}}, \bibinfo {author} {\bibfnamefont {S.}~\bibnamefont {Takayoshi}},
  \bibinfo {author} {\bibfnamefont {T.}~\bibnamefont {Kaneko}}, \bibinfo
  {author} {\bibfnamefont {Z.}~\bibnamefont {Sun}}, \bibinfo {author}
  {\bibfnamefont {D.}~\bibnamefont {Golez}}, \bibinfo {author} {\bibfnamefont
  {A.~J.}\ \bibnamefont {Millis}},\ and\ \bibinfo {author} {\bibfnamefont
  {P.}~\bibnamefont {Werner}},\ }\bibfield  {title} {\bibinfo {title}
  {Exploring nonequilibrium phases of photo-doped mott insulators with
  generalized gibbs ensembles},\ }\href
  {https://doi.org/10.1038/s42005-021-00799-7} {\bibfield  {journal} {\bibinfo
  {journal} {Communications Physics}\ }\textbf {\bibinfo {volume} {5}},\
  \bibinfo {pages} {23} (\bibinfo {year} {2022})}\BibitemShut {NoStop}%
\bibitem [{\citenamefont {Lenarcic}\ and\ \citenamefont
  {Prelovsek}(2013)}]{le.pr.13}%
  \BibitemOpen
  \bibfield  {author} {\bibinfo {author} {\bibfnamefont {Z.}~\bibnamefont
  {Lenarcic}}\ and\ \bibinfo {author} {\bibfnamefont {P.}~\bibnamefont
  {Prelovsek}},\ }\bibfield  {title} {\bibinfo {title} {Ultrafast charge
  recombination in a photoexcited mott-hubbard insulator},\ }\href
  {https://doi.org/10.1103/PhysRevLett.111.016401} {\bibfield  {journal}
  {\bibinfo  {journal} {Phys. Rev. Lett.}\ }\textbf {\bibinfo {volume} {111}},\
  \bibinfo {pages} {016401} (\bibinfo {year} {2013})}\BibitemShut {NoStop}%
\bibitem [{\citenamefont {K{\"u}nzel}\ \emph {et~al.}(2024)\citenamefont
  {K{\"u}nzel}, \citenamefont {Erpenbeck}, \citenamefont {Werner},
  \citenamefont {Arrigoni}, \citenamefont {Gull}, \citenamefont {Cohen},\ and\
  \citenamefont {Eckstein}}]{ku.er.24}%
  \BibitemOpen
  \bibfield  {author} {\bibinfo {author} {\bibfnamefont {F.}~\bibnamefont
  {K{\"u}nzel}}, \bibinfo {author} {\bibfnamefont {A.}~\bibnamefont
  {Erpenbeck}}, \bibinfo {author} {\bibfnamefont {D.}~\bibnamefont {Werner}},
  \bibinfo {author} {\bibfnamefont {E.}~\bibnamefont {Arrigoni}}, \bibinfo
  {author} {\bibfnamefont {E.}~\bibnamefont {Gull}}, \bibinfo {author}
  {\bibfnamefont {G.}~\bibnamefont {Cohen}},\ and\ \bibinfo {author}
  {\bibfnamefont {M.}~\bibnamefont {Eckstein}},\ }\bibfield  {title} {\bibinfo
  {title} {Numerically exact simulation of photodoped mott insulators},\ }\href
  {https://doi.org/10.1103/PhysRevLett.132.176501} {\bibfield  {journal}
  {\bibinfo  {journal} {Phys. Rev. Lett.}\ }\textbf {\bibinfo {volume} {132}},\
  \bibinfo {pages} {176501} (\bibinfo {year} {2024})}\BibitemShut {NoStop}%
\bibitem [{\citenamefont {Picano}\ \emph {et~al.}(2021)\citenamefont {Picano},
  \citenamefont {Li},\ and\ \citenamefont {Eckstein}}]{pi.li.21}%
  \BibitemOpen
  \bibfield  {author} {\bibinfo {author} {\bibfnamefont {A.}~\bibnamefont
  {Picano}}, \bibinfo {author} {\bibfnamefont {J.}~\bibnamefont {Li}},\ and\
  \bibinfo {author} {\bibfnamefont {M.}~\bibnamefont {Eckstein}},\ }\bibfield
  {title} {\bibinfo {title} {Quantum boltzmann equation for strongly correlated
  electrons},\ }\href@noop {} {\bibfield  {journal} {\bibinfo  {journal} {Phys.
  Rev. B}\ }\textbf {\bibinfo {volume} {104}},\ \bibinfo {pages} {085108}
  (\bibinfo {year} {2021})}\BibitemShut {NoStop}%
\bibitem [{\citenamefont {Dorda}\ \emph {et~al.}(2017)\citenamefont {Dorda},
  \citenamefont {Sorantin}, \citenamefont {von~der Linden},\ and\ \citenamefont
  {Arrigoni}}]{do.so.17}%
  \BibitemOpen
  \bibfield  {author} {\bibinfo {author} {\bibfnamefont {A.}~\bibnamefont
  {Dorda}}, \bibinfo {author} {\bibfnamefont {M.}~\bibnamefont {Sorantin}},
  \bibinfo {author} {\bibfnamefont {W.}~\bibnamefont {von~der Linden}},\ and\
  \bibinfo {author} {\bibfnamefont {E.}~\bibnamefont {Arrigoni}},\ }\bibfield
  {title} {\bibinfo {title} {Optimized auxiliary representation of
  non-markovian impurity problems by a lindblad equation},\ }\href
  {https://doi.org/10.1088/1367-2630/aa6ccc} {\bibfield  {journal} {\bibinfo
  {journal} {New J. Phys.}\ }\textbf {\bibinfo {volume} {19}},\ \bibinfo
  {pages} {063005} (\bibinfo {year} {2017})}\BibitemShut {NoStop}%
\bibitem [{\citenamefont {Chen}\ \emph {et~al.}(2019)\citenamefont {Chen},
  \citenamefont {Cohen},\ and\ \citenamefont {Galperin}}]{ch.co.19}%
  \BibitemOpen
  \bibfield  {author} {\bibinfo {author} {\bibfnamefont {F.}~\bibnamefont
  {Chen}}, \bibinfo {author} {\bibfnamefont {G.}~\bibnamefont {Cohen}},\ and\
  \bibinfo {author} {\bibfnamefont {M.}~\bibnamefont {Galperin}},\ }\bibfield
  {title} {\bibinfo {title} {Auxiliary master equation for nonequilibrium
  dual-fermion approach},\ }\href
  {https://doi.org/10.1103/PhysRevLett.122.186803} {\bibfield  {journal}
  {\bibinfo  {journal} {Phys. Rev. Lett.}\ }\textbf {\bibinfo {volume} {122}},\
  \bibinfo {pages} {186803} (\bibinfo {year} {2019})}\BibitemShut {NoStop}%
\bibitem [{\citenamefont {Haug}\ and\ \citenamefont {Jauho}(1998)}]{ha.ja}%
  \BibitemOpen
  \bibfield  {author} {\bibinfo {author} {\bibfnamefont {H.}~\bibnamefont
  {Haug}}\ and\ \bibinfo {author} {\bibfnamefont {A.-P.}\ \bibnamefont
  {Jauho}},\ }\href {http://www.springer.com/us/book/9783540735618} {\emph
  {\bibinfo {title} {Quantum Kinetics in Transport and Optics of
  Semiconductors}}}\ (\bibinfo  {publisher} {Springer},\ \bibinfo {address}
  {Heidelberg},\ \bibinfo {year} {1998})\BibitemShut {NoStop}%
\bibitem [{\citenamefont {Schwinger}(1961)}]{schw.61}%
  \BibitemOpen
  \bibfield  {author} {\bibinfo {author} {\bibfnamefont {J.}~\bibnamefont
  {Schwinger}},\ }\bibfield  {title} {\bibinfo {title} {Brownian motion of a
  quantum oscillator},\ }\href@noop {} {\bibfield  {journal} {\bibinfo
  {journal} {J. Math. Phys.}\ }\textbf {\bibinfo {volume} {2}},\ \bibinfo
  {pages} {407} (\bibinfo {year} {1961})}\BibitemShut {NoStop}%
\bibitem [{\citenamefont {Keldysh}(1965)}]{keld.65}%
  \BibitemOpen
  \bibfield  {author} {\bibinfo {author} {\bibfnamefont {L.~V.}\ \bibnamefont
  {Keldysh}},\ }\bibfield  {title} {\bibinfo {title} {Diagram technique for
  nonequilibrium processes},\ }\href@noop {} {\bibfield  {journal} {\bibinfo
  {journal} {Sov. Phys. JETP}\ }\textbf {\bibinfo {volume} {20}},\ \bibinfo
  {pages} {1018} (\bibinfo {year} {1965})}\BibitemShut {NoStop}%
\bibitem [{\citenamefont {Kadanoff}\ and\ \citenamefont
  {Baym}(1962)}]{kad.baym}%
  \BibitemOpen
  \bibfield  {author} {\bibinfo {author} {\bibfnamefont {L.~P.}\ \bibnamefont
  {Kadanoff}}\ and\ \bibinfo {author} {\bibfnamefont {G.}~\bibnamefont
  {Baym}},\ }\href@noop {} {\emph {\bibinfo {title} {Quantum Statistical
  Mechanics: Green's Function Methods in Equilibrium and Nonequilibrium
  Problems}}}\ (\bibinfo  {publisher} {Addison-Wesley},\ \bibinfo {address}
  {Redwood City, CA},\ \bibinfo {year} {1962})\BibitemShut {NoStop}%
\bibitem [{\citenamefont {Rammer}\ and\ \citenamefont
  {Smith}(1986)}]{ra.sm.86}%
  \BibitemOpen
  \bibfield  {author} {\bibinfo {author} {\bibfnamefont {J.}~\bibnamefont
  {Rammer}}\ and\ \bibinfo {author} {\bibfnamefont {H.}~\bibnamefont {Smith}},\
  }\bibfield  {title} {\bibinfo {title} {Quantum field-theoretical methods in
  transport theory of metals},\ }\href
  {https://doi.org/10.1103/RevModPhys.58.323} {\bibfield  {journal} {\bibinfo
  {journal} {Rev. Mod. Phys.}\ }\textbf {\bibinfo {volume} {58}},\ \bibinfo
  {pages} {323} (\bibinfo {year} {1986})}\BibitemShut {NoStop}%
\bibitem [{Note1()}]{Note1}%
  \BibitemOpen
  \bibinfo {note} {Here and in the following, we use underscore $\protect
  \underline {X}$ to denote this Keldysh structure containing the retarded
  $X^R$, Keldysh $X^K$ and advanced $X^A$ components~\cite {ha.ja,ra.sm.86}.
  $X$ can be any two-point function, such as $\Delta $, $G$, or $\Sigma $, or
  their differences, such as the quantities $\varepsilon $ or $\mu $ below. In
  addition, we shall omit the argument $\omega $ unless
  neccessary.}\BibitemShut {Stop}%
\bibitem [{\citenamefont {Werner}\ \emph {et~al.}(2023)\citenamefont {Werner},
  \citenamefont {Lotze},\ and\ \citenamefont {Arrigoni}}]{we.lo.23}%
  \BibitemOpen
  \bibfield  {author} {\bibinfo {author} {\bibfnamefont {D.}~\bibnamefont
  {Werner}}, \bibinfo {author} {\bibfnamefont {J.}~\bibnamefont {Lotze}},\ and\
  \bibinfo {author} {\bibfnamefont {E.}~\bibnamefont {Arrigoni}},\ }\bibfield
  {title} {\bibinfo {title} {Configuration interaction based nonequilibrium
  steady state impurity solver},\ }\href
  {https://doi.org/10.1103/PhysRevB.107.075119} {\bibfield  {journal} {\bibinfo
   {journal} {Phys. Rev. B}\ }\textbf {\bibinfo {volume} {107}},\ \bibinfo
  {pages} {075119} (\bibinfo {year} {2023})}\BibitemShut {NoStop}%
\bibitem [{\citenamefont {Dorda}\ \emph {et~al.}(2015)\citenamefont {Dorda},
  \citenamefont {Ganahl}, \citenamefont {Evertz}, \citenamefont {von~der
  Linden},\ and\ \citenamefont {Arrigoni}}]{do.ga.15}%
  \BibitemOpen
  \bibfield  {author} {\bibinfo {author} {\bibfnamefont {A.}~\bibnamefont
  {Dorda}}, \bibinfo {author} {\bibfnamefont {M.}~\bibnamefont {Ganahl}},
  \bibinfo {author} {\bibfnamefont {H.~G.}\ \bibnamefont {Evertz}}, \bibinfo
  {author} {\bibfnamefont {W.}~\bibnamefont {von~der Linden}},\ and\ \bibinfo
  {author} {\bibfnamefont {E.}~\bibnamefont {Arrigoni}},\ }\bibfield  {title}
  {\bibinfo {title} {Auxiliary master equation approach within matrix product
  states: Spectral properties of the nonequilibrium anderson impurity model},\
  }\href {https://doi.org/10.1103/PhysRevB.92.125145} {\bibfield  {journal}
  {\bibinfo  {journal} {Phys. Rev. B}\ }\textbf {\bibinfo {volume} {92}},\
  \bibinfo {pages} {125145} (\bibinfo {year} {2015})}\BibitemShut {NoStop}%
\bibitem [{\citenamefont {Sorantin}\ \emph {et~al.}(2019)\citenamefont
  {Sorantin}, \citenamefont {Fugger}, \citenamefont {Dorda}, \citenamefont
  {von~der Linden},\ and\ \citenamefont {Arrigoni}}]{so.fu.19}%
  \BibitemOpen
  \bibfield  {author} {\bibinfo {author} {\bibfnamefont {M.~E.}\ \bibnamefont
  {Sorantin}}, \bibinfo {author} {\bibfnamefont {D.~M.}\ \bibnamefont
  {Fugger}}, \bibinfo {author} {\bibfnamefont {A.}~\bibnamefont {Dorda}},
  \bibinfo {author} {\bibfnamefont {W.}~\bibnamefont {von~der Linden}},\ and\
  \bibinfo {author} {\bibfnamefont {E.}~\bibnamefont {Arrigoni}},\ }\bibfield
  {title} {\bibinfo {title} {Auxiliary master equation approach within
  stochastic wave functions: Application to the interacting resonant level
  model},\ }\href {https://doi.org/10.1103/PhysRevE.99.043303} {\bibfield
  {journal} {\bibinfo  {journal} {Phys. Rev. E}\ }\textbf {\bibinfo {volume}
  {99}},\ \bibinfo {pages} {043303} (\bibinfo {year} {2019})}\BibitemShut
  {NoStop}%
\bibitem [{Note2()}]{Note2}%
  \BibitemOpen
  \bibinfo {note} {Combined by a configuraion interaction (CI) approach~\cite
  {we.lo.23}}\BibitemShut {NoStop}%
\bibitem [{\citenamefont {Keiter}\ and\ \citenamefont
  {Kimball}(1970)}]{ke.ki.70}%
  \BibitemOpen
  \bibfield  {author} {\bibinfo {author} {\bibfnamefont {H.}~\bibnamefont
  {Keiter}}\ and\ \bibinfo {author} {\bibfnamefont {J.~C.}\ \bibnamefont
  {Kimball}},\ }\bibfield  {title} {\bibinfo {title} {Perturbation technique
  for the anderson hamiltonian},\ }\href
  {https://doi.org/10.1103/PhysRevLett.25.672} {\bibfield  {journal} {\bibinfo
  {journal} {Phys. Rev. Lett.}\ }\textbf {\bibinfo {volume} {25}},\ \bibinfo
  {pages} {672} (\bibinfo {year} {1970})}\BibitemShut {NoStop}%
\bibitem [{\citenamefont {Coleman}(1984)}]{cole.84}%
  \BibitemOpen
  \bibfield  {author} {\bibinfo {author} {\bibfnamefont {P.}~\bibnamefont
  {Coleman}},\ }\href@noop {} {\bibfield  {journal} {\bibinfo  {journal} {Phys.
  Rev. B}\ }\textbf {\bibinfo {volume} {29}},\ \bibinfo {pages} {3035}
  (\bibinfo {year} {1984})}\BibitemShut {NoStop}%
\bibitem [{\citenamefont {Eckstein}\ and\ \citenamefont
  {Werner}(2010)}]{ec.we.10}%
  \BibitemOpen
  \bibfield  {author} {\bibinfo {author} {\bibfnamefont {M.}~\bibnamefont
  {Eckstein}}\ and\ \bibinfo {author} {\bibfnamefont {P.}~\bibnamefont
  {Werner}},\ }\bibfield  {title} {\bibinfo {title} {Nonequilibrium dynamical
  mean-field calculations based on the noncrossing approximation and its
  generalizations},\ }\href {https://doi.org/10.1103/PhysRevB.82.115115}
  {\bibfield  {journal} {\bibinfo  {journal} {Phys. Rev. B}\ }\textbf {\bibinfo
  {volume} {82}},\ \bibinfo {pages} {115115} (\bibinfo {year}
  {2010})}\BibitemShut {NoStop}%
\bibitem [{\citenamefont {Cohen}\ \emph {et~al.}(2015)\citenamefont {Cohen},
  \citenamefont {Gull}, \citenamefont {Reichman},\ and\ \citenamefont
  {Millis}}]{co.gu.15}%
  \BibitemOpen
  \bibfield  {author} {\bibinfo {author} {\bibfnamefont {G.}~\bibnamefont
  {Cohen}}, \bibinfo {author} {\bibfnamefont {E.}~\bibnamefont {Gull}},
  \bibinfo {author} {\bibfnamefont {D.~R.}\ \bibnamefont {Reichman}},\ and\
  \bibinfo {author} {\bibfnamefont {A.~J.}\ \bibnamefont {Millis}},\ }\bibfield
   {title} {\bibinfo {title} {Taming the dynamical sign problem in real-time
  evolution of quantum many-body problems},\ }\href
  {https://doi.org/10.1103/PhysRevLett.115.266802} {\bibfield  {journal}
  {\bibinfo  {journal} {Phys. Rev. Lett.}\ }\textbf {\bibinfo {volume} {115}},\
  \bibinfo {pages} {266802} (\bibinfo {year} {2015})}\BibitemShut {NoStop}%
\bibitem [{\citenamefont {N\'u\~nez Fern\'andez}\ \emph
  {et~al.}(2022)\citenamefont {N\'u\~nez Fern\'andez}, \citenamefont {Jeannin},
  \citenamefont {Dumitrescu}, \citenamefont {Kloss}, \citenamefont {Kaye},
  \citenamefont {Parcollet},\ and\ \citenamefont {Waintal}}]{nu.je.22}%
  \BibitemOpen
  \bibfield  {author} {\bibinfo {author} {\bibfnamefont {Y.}~\bibnamefont
  {N\'u\~nez Fern\'andez}}, \bibinfo {author} {\bibfnamefont {M.}~\bibnamefont
  {Jeannin}}, \bibinfo {author} {\bibfnamefont {P.~T.}\ \bibnamefont
  {Dumitrescu}}, \bibinfo {author} {\bibfnamefont {T.}~\bibnamefont {Kloss}},
  \bibinfo {author} {\bibfnamefont {J.}~\bibnamefont {Kaye}}, \bibinfo {author}
  {\bibfnamefont {O.}~\bibnamefont {Parcollet}},\ and\ \bibinfo {author}
  {\bibfnamefont {X.}~\bibnamefont {Waintal}},\ }\bibfield  {title} {\bibinfo
  {title} {Learning feynman diagrams with tensor trains},\ }\href
  {https://doi.org/10.1103/PhysRevX.12.041018} {\bibfield  {journal} {\bibinfo
  {journal} {Phys. Rev. X}\ }\textbf {\bibinfo {volume} {12}},\ \bibinfo
  {pages} {041018} (\bibinfo {year} {2022})}\BibitemShut {NoStop}%
\bibitem [{\citenamefont {Eckstein}(2024)}]{ecks.24u}%
  \BibitemOpen
  \bibfield  {author} {\bibinfo {author} {\bibfnamefont {M.}~\bibnamefont
  {Eckstein}},\ }\bibfield  {title} {\bibinfo {title} {Solving quantum impurity
  models in the non -- equilibrium steady state with tensor trains}} (\bibinfo
  {year} {2024}),\ \bibinfo {note} {arXiv:2410.19707}\BibitemShut {NoStop}%
\bibitem [{\citenamefont {Chitambar}\ and\ \citenamefont
  {Gour}(2019)}]{ch.go.19}%
  \BibitemOpen
  \bibfield  {author} {\bibinfo {author} {\bibfnamefont {E.}~\bibnamefont
  {Chitambar}}\ and\ \bibinfo {author} {\bibfnamefont {G.}~\bibnamefont
  {Gour}},\ }\bibfield  {title} {\bibinfo {title} {Quantum resource theories},\
  }\href {https://doi.org/10.1103/RevModPhys.91.025001} {\bibfield  {journal}
  {\bibinfo  {journal} {Rev. Mod. Phys.}\ }\textbf {\bibinfo {volume} {91}},\
  \bibinfo {pages} {025001} (\bibinfo {year} {2019})}\BibitemShut {NoStop}%
\bibitem [{Note3()}]{Note3}%
  \BibitemOpen
  \bibinfo {note} {In Ref.~\cite {ch.go.19} the auxiliary system is referred to
  as reference system.}\BibitemShut {Stop}%
\bibitem [{Note4()}]{Note4}%
  \BibitemOpen
  \bibinfo {note} {$k$ can, in principle, change as a function of
  $n$.}\BibitemShut {Stop}%
\bibitem [{Note5()}]{Note5}%
  \BibitemOpen
  \bibinfo {note} {The condition number is a measure of how ill-conditioned the
  matrix inversion is, see. e.g. \protect \url
  {https://numpy.org/doc/2.1/reference/generated/numpy.linalg.cond.html}}\BibitemShut
  {NoStop}%
\bibitem [{Note6()}]{Note6}%
  \BibitemOpen
  \bibinfo {note} {Notice that ${\protect \underline {\Delta }_\protect \text
  {des}}_m$ may become unphysical, for example non-causal. This is, however,
  not an issue, as the fitting procedure ensures that ${\protect \underline
  {\Delta }_\protect \text {aux}}_m$ remains physical.}\BibitemShut {Stop}%
\bibitem [{Note7()}]{Note7}%
  \BibitemOpen
  \bibinfo {note} {By ``plain'' we mean without the FI improvement,
  correpsonding to $N_\protect \text {T}=1$.}\BibitemShut {Stop}%
\bibitem [{sup()}]{suppmat}%
  \BibitemOpen
  \bibinfo {note} {See supplemental Material at [URL will be inserted by the
  publisher] for details.}\BibitemShut {Stop}%
\bibitem [{Note8()}]{Note8}%
  \BibitemOpen
  \bibinfo {note} {The edges of the DOS are smoothed with a ``fictitious''
  temperature $0.5\Gamma $.}\BibitemShut {Stop}%
\bibitem [{Note9()}]{Note9}%
  \BibitemOpen
  \bibinfo {note} {We omitted the Keldysh component, since it does not give any
  additional insight.}\BibitemShut {Stop}%
\bibitem [{Note10()}]{Note10}%
  \BibitemOpen
  \bibinfo {note} {The weight in the fit function is set to $10$ for $|\omega
  |/\Gamma < 2$ and to $1$ otherwise.}\BibitemShut {Stop}%
\end{thebibliography}%

\appendix

\section{Supplemental material}

Since we have a certain freedom in choosing the details of the procedure, such as in the choice of $k$ or whether to use the deterministic or the random procedure
described in the main text,
it is convenient to first test different options in order to find out the most convenient one. We do this by applying the approach for the  numerically ``cheap'' case of an auxiliary bath with $N_\text{b}=4$ and compare the self-energy obtained in this way with a reference evaluated with ``plain'' $N_\text{b}=8$ which is expected to be more accurate even without this FI procedure.

Unless otherwise specified, we consider a bath with a flat DOS of
width $10 \Gamma$
with the following parameters
$U/\Gamma = 6$, $-\IIm \Delta^\text{R}(\omega = 0)/\Gamma = 1$, $\varepsilon_{\text{imp}} = -U/2$, $T/\Gamma = 0.05$. Moreover we denote
$N_\text{T}$ as the maximum number
of auxiliary systems included in the sum \eqref{sigmaimp}, corresponding to a set of deviations
\[
	\{\meps_1, \meps_2,\cdots,\meps_{N_\text{T}}\}.
\] Similarly, we denote by
$\mmu_{n_{\max}}$
the corresponding optimized deviations \eqref{mu}
for $n_{\max}=1,\cdots,N_\text{T}$.

As mentioned in
the results section
we compare self energies with $N_\text{b}=4$ and different versions of the FI approach with the ones
obtained with ``plain'' eight bath sites ($N_\text{b} = 8$) (i.e. no FI procedure).
Since the ``plain'' $N_\text{b} = 8$ results are expected to be very accurate, this provides an error estimate for the $N_\text{b}=4$ FI calculations as
\newcommand{\error}{\chi_{\Sigma}}
\begin{equation}
	\label{error}
	\begin{split}
		\error =& \frac{1}{2 \omega_\text{max}} \int_{\omega = -\omega_\text{max}}^{\omega = \omega_\text{max}} \Bigl(|\IIm(\Sigma^\text{R}_{N_\text{b} = 8}) - \IIm(\Sigma^\text{R}_{N_\text{b} = 4,\text{FI}})|^2 \\
		&+ |\IIm(\Sigma^\text{K}_{N_\text{b} = 8}) - \IIm(\Sigma^\text{K}_{N_\text{b} = 4,\text{FI}})|^2\Bigr) d\omega.
	\end{split}
\end{equation}

\begin{figure}
	\centering
	\includegraphics[scale=0.4]{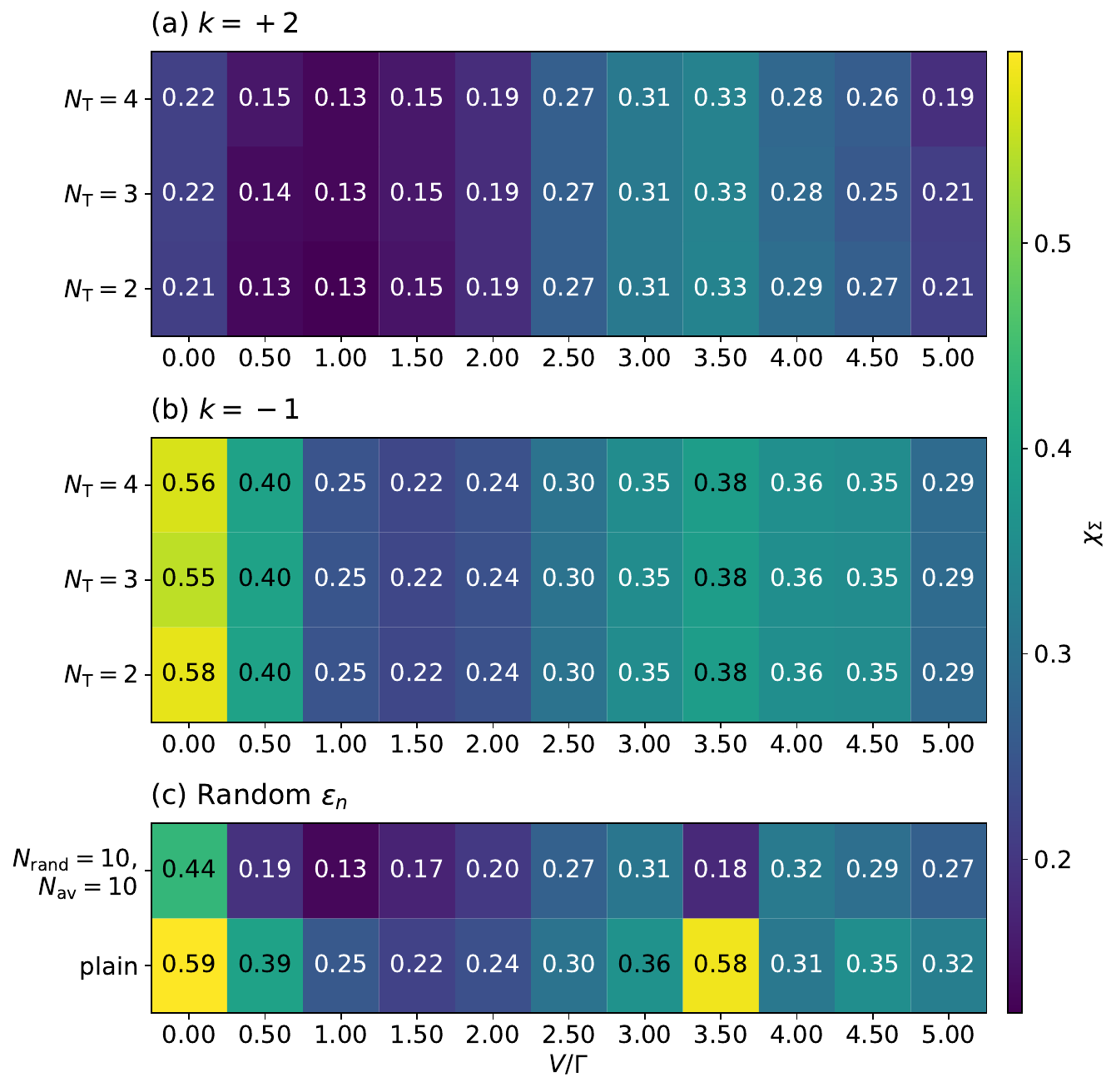}
	\caption{Error $\error$ (Eq.~\ref{error}) of results obtained using four bath sites ($N_\text{b} = 4$) in combination with the FI based approach
		at $T/\Gamma = 0.05$ and voltage drop ranging from $V/\Gamma = 0$ to $5$ in steps of $0.5$. (a) and (b) use the deterministic algorithm
		(see text)
		and (c) uses the ``random'' one.
		In (a) we use $k=2$, while in (b) we use $k=-1$.
	}
	\label{fig:err}
\end{figure}

For the deterministic algorithm described in
the text
we try the factors $k = -1$ and $k = 2$. Table \ref{fig:err}(a) shows that using $k = -1$ the FI approach gives no improvement, whereas from table \ref{fig:err}(b) it becomes evident, that $k = +2$ produces a strong reduction of $\error$ over the full voltage range.
On the other hand, we notice that increasing $N_\text{T}$ beyond $2$ hardly produce any improvement.
This is an important result which suggests that in fact it is sufficient to solve for two impurity solvers in order to substantially improve the accuracy of the calculation. Of course this is valid for the present model with a flat band. The story may be different for a more subtle density of states. This will be the goal of future investigations.

It is, nevertheless, important to understand the reasons behind this behavior.

\begin{figure}
	\centering
	\includegraphics[scale=0.27]{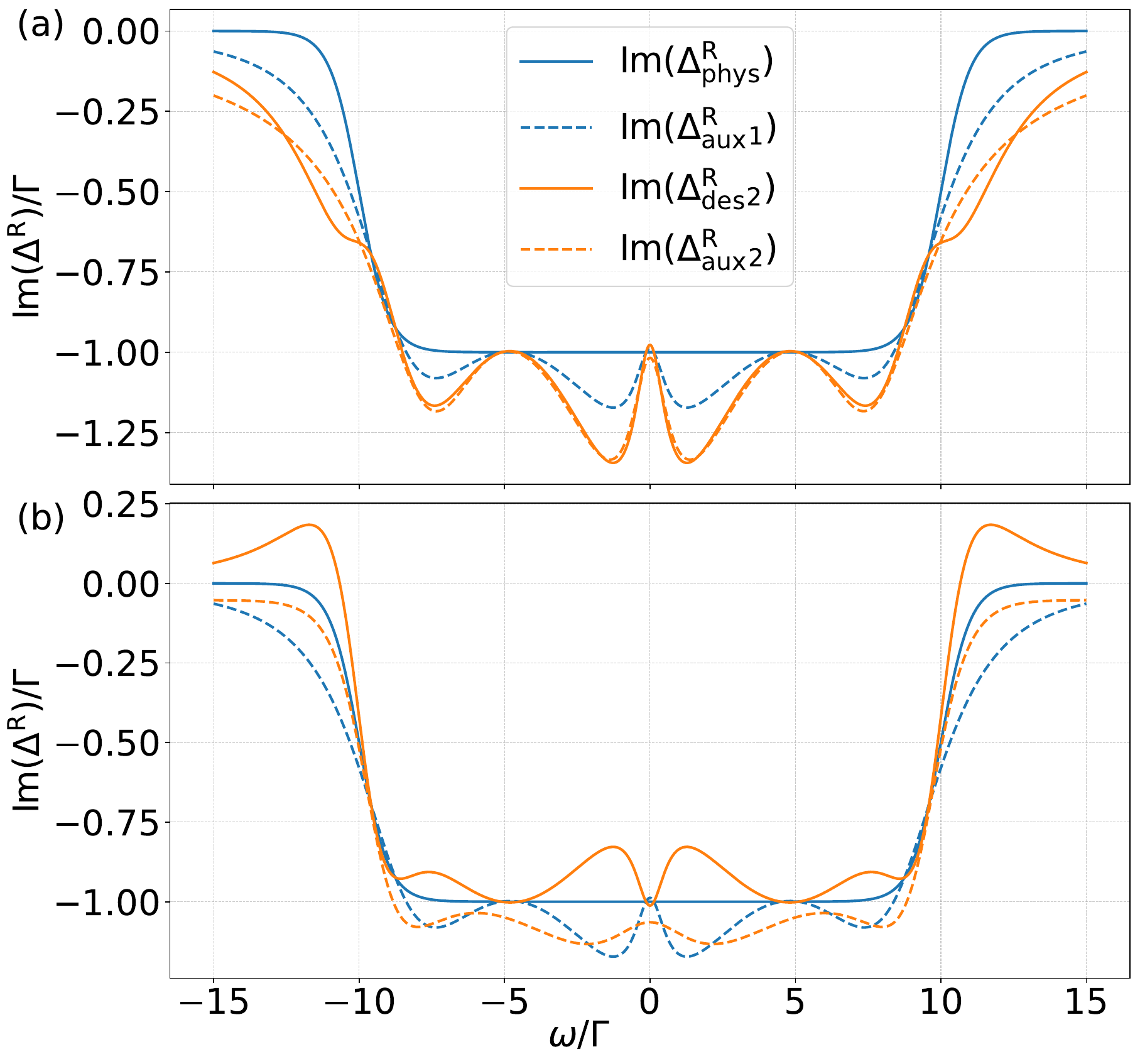}
	\caption{$\IIm({\Delta^\text{R}\aux}_2)$ (orange, dashed) fitted to the corresponding
		``desired'' $\IIm({\Delta^\text{R}\des}_2)$
		(orange, solid) with $k = 2$ (a) and $k = -1$ (b). For $k=2$ the fit is extremely accurate, in contrast to $k=-1$. Both curves also show
		$\IIm({\Delta^\text{R}\aux}_1)$ and $\IIm({{\Delta^\text{R}_\text{phys}}})$ for comparison
	}
	\label{fig:Delta_R_taylor}
\end{figure}

First, we want to understand, why $k=2$ is more accurate than $k=-1$.
$k=2$ means that the $m=2$ hybridization ${\matK{\Delta}\des}_2$ is farther away from the pyhsical one than the $m=1$ one. On the other hand, $k=-1$ corresponds to a ${\matK{\Delta}\des}_2$ on the ``other side'' of the physical one.
So, why is it more convenient to go ``further away'' from the correct hybridization rather than to go ``to the other side''?
In order to address this, we look at the ``desired''
hybridization functions (${\matK{\Delta}\des}_2 =\matK{\Delta}\phys + k \matK{\mu}_1$) and the corresponding fitted ones
(${\matK{\Delta}\aux}_2$) for
$N_\text{T} = 2$.
The imaginary part of their retarded component are shown in figure \ref{fig:Delta_R_taylor} (a) for $k = +2$ and in \ref{fig:Delta_R_taylor} (b) for $k = -1$~\footnote{We omitted the Keldysh component, since it does not give any additional insight.} and compared with
the physical hybridization
${\matK{\Delta}\phys} (= {\matK{\Delta}_\text{des}}_1)$
and the first auxiliary one
${\matK{\Delta}\aux}_1$.
The figures clearly show that the fit
to ${\matK{\Delta}\des}_2$
is far more accurate for $k=+2$ than for $k=-1$.
The reason for this better fit is due to the fact that the desired hybridization
function for $k=+2$
simply amplifies the features of the first fit (${\matK{\Delta}\aux}_1$).

\begin{figure}
	\centering
	\includegraphics[scale=0.3]{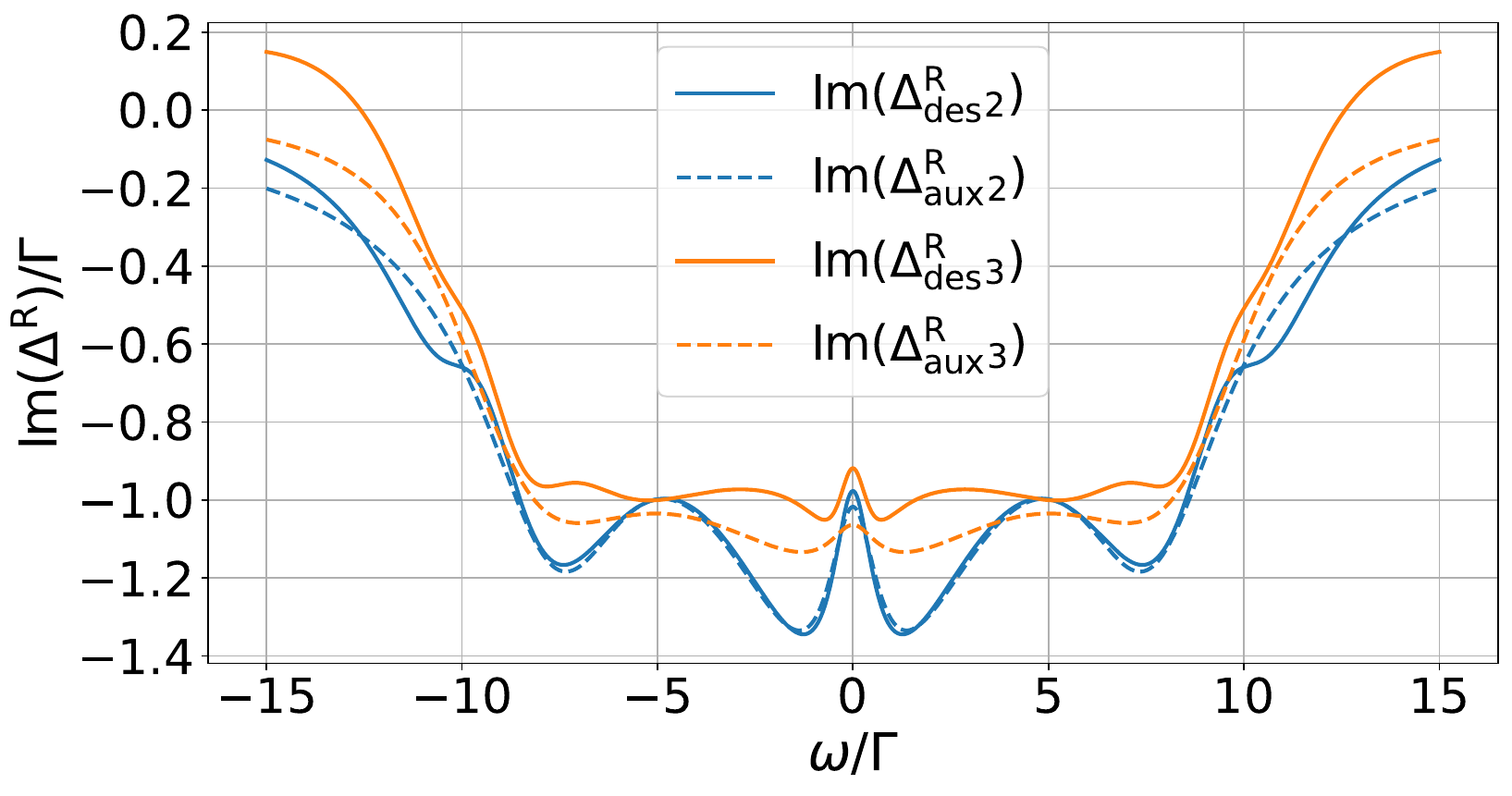}
	\caption{
		$\IIm({\Delta^\text{R}\des}_{m})$ (solid) as well as its fit
		$\IIm({\Delta^\text{R}\aux}_{m})$ (dashed) for $m=2$ (blue) and $3$ (orange).
		While the fit for $m=2$
		shows almost perfect coincidence, the one for $m=3$ is quite poor, thus providing no improvement in the FI procedure.
	}
	\label{fig:Delta_R_taylor_more_plus}
\end{figure}

We now address
the question why including more than two $\matK{\Delta}\aux$ in \eqref{sigmaimp}
doesn't improve the accuracy of the self-energy.
Similarly to the previous analysis,
in Fig.~\ref{fig:Delta_R_taylor_more_plus}
we plot the ``desired''
${\matK{\Delta}\des}_{m}=\matK{\Delta}\phys + k \matK{\mu}_{m-1}$ (solid line) along with its corresponding fit
${\matK{\Delta}\aux}_{m}$
(dashed line with same color). We restrict to the imaginary part of the retarded component.

We fix $k=2$, as suggested by the previous analysis and we show the results for $m = 2$ and $m = 3$ with
$N_\text{T}=3$.
As one can see, while the fit to the ${\Delta^\text{R}_\text{des}}_2$ curve is excellent, the one to the ${\Delta^\text{R}_\text{des}}_3$ curve is quite poor.
As a consequence, inclusion of ${\matK{\Delta}\aux}_3$ and corresponding $\meps_3$ (cf. \eqref{epsn})
does not make $\mmu$ \eqref{mu} smaller and, consequently, does not produce a more accurate self-energy (\eqref{sigmaimp}).

\begin{figure}
	\centering
	\includegraphics[scale=0.325]{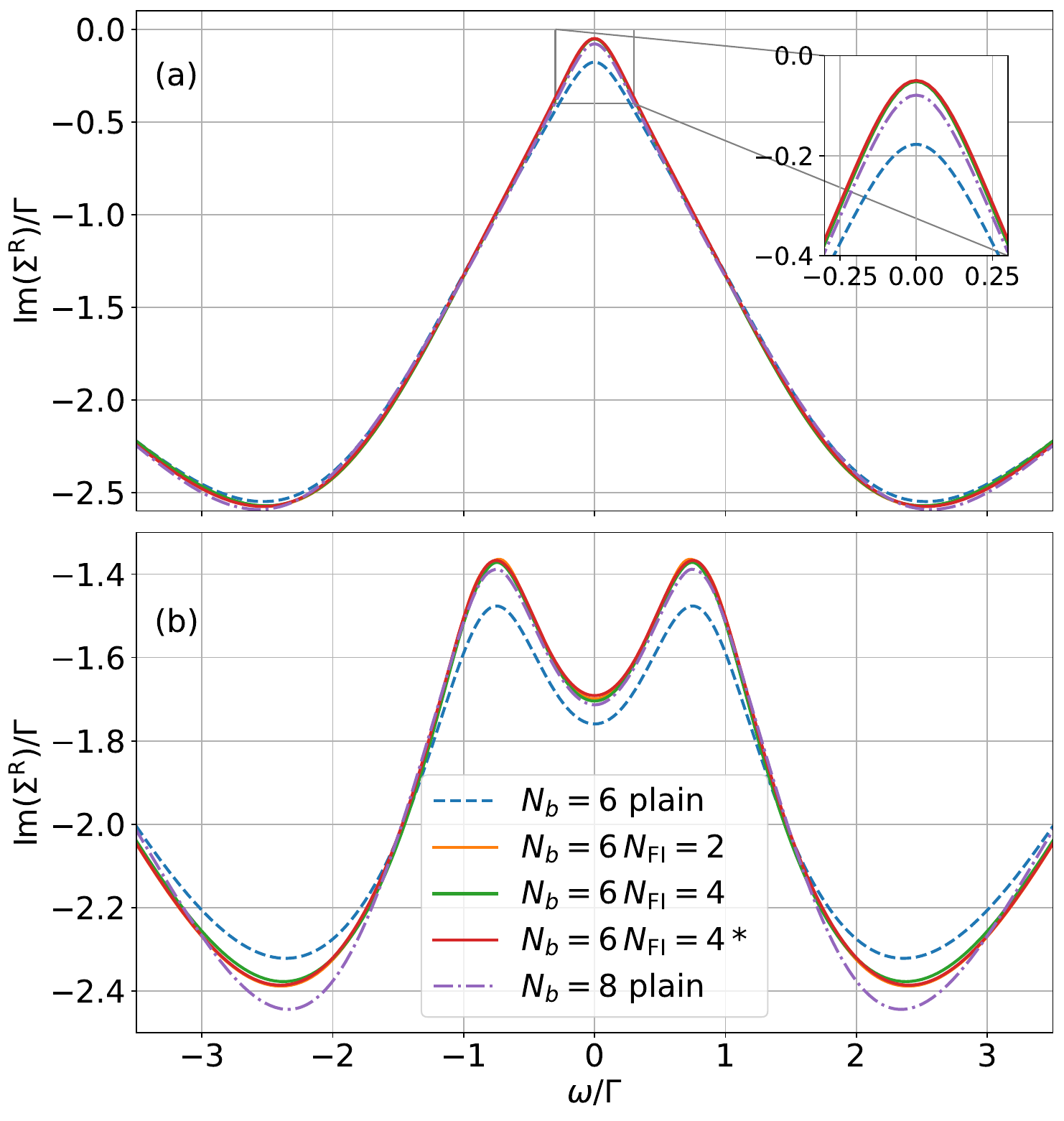}
	\caption{Comparison of plain calculations with those obtained via the FI approach for $V/\Gamma = 0$ (a) and $V/\Gamma = 1.5$ (b). $\IIm(\Sigma^\text{R})$ for $N_\text{b} = 6$ does not change significantly when increasing the number of included hybridization functions. The $N_\text{T} = 4*$ results have been obtained by combining two $N_\text{T} = 2$ sets of hybridization functions with
		weights in the fit (see text).}
	\label{fig:SE_Nb6_Taylor}
\end{figure}

This may change when using more bath sites $N_\text{b}>4$,
since one can better reproduce subtle features.
To investigate this,
we show in figure~\ref{fig:SE_Nb6_Taylor} results for $N_\text{b} = 6$ at $T/\Gamma=0.02$ and $V/\Gamma = 0$ (a) and $V/\Gamma = 1.5$ (b) respectively. Here we compare the self-energy obtained with $N_\text{T}=2$ with the one obtained with two additional steps ($N_\text{T}=4$) of the deterministic algorithm.
This figure also includes a plot (termed $N_\text{T}=4*$)
where we combined two $N_\text{T} = 2$ sets of hybridization functions, where
the second set
used weights~\footnote{The weight in the fit function
	is set to $10$ for $|\omega|/\Gamma < 2$ and to $1$ otherwise.} in the fit of the hybridization functions.
Comparison with the benchmark ($N_\text{b} = 8$) shows that even in this case there is no improvement upon increasing $N_\text{T}$.
Therefore, we will use $N_\text{T} = 2$ for the deterministic algorithm.
The results also show a significant improvement with respect to the plain AMEA result, thus emphasizing the importance of using the FI approach even by evaluating just one more hybridization function.

To test the ``random'' algorithm
we choose

\begin{equation}
	\IIm {\Delta^\text{R/K}_\text{des}}_m = \IIm \Delta^\text{R/K}_\text{phys} + f_\text{rand}^\text{R/K}(\omega) \ \ \ \forall \ m > 1
\end{equation}

with $f_\text{rand}^\text{R/K}(\omega)$ having five equidistant points between $-\omega_\text{max}$ and $\omega_\text{max}$, where at each point a random value between $-0.3$ and $0.3$ is taken and the space in between is smoothly interpolated. Figure~\ref{fig:err} (c) shows the error $\error$ in the self-energy for this algorithm.
We used $N_\text{T} = 10$ and $N_\text{av} = 10$. The results show a clear improvement with respect to the plain case, but mostly less than the one obtained by the deterministic algorithm with $k = +2$, (cf. Fig.~\ref{fig:err}(a)) which is also computationally much less costly. For this reason, in this paper we adopt the deterministic algorithm with $k=2$.

\end{document}